\documentclass[showkeys,superscriptaddress,floatfix,prd,10pt,aps]{revtex4-2}
\usepackage{graphicx,epstopdf}
\usepackage[caption=false]{subfig}
\usepackage{appendix}
\usepackage[T1]{fontenc}
\usepackage{lmodern}
\usepackage[dvipsnames,x11names]{xcolor}
\usepackage[colorlinks=true,linkcolor=NavyBlue,citecolor=ForestGreen,urlcolor=NavyBlue]{hyperref}
\usepackage[sort&compress]{natbib}
\usepackage{lipsum}
\usepackage{morefloats}
\usepackage[pdf]{pstricks}
\usepackage{amsmath}
\usepackage{amssymb}
\usepackage{amsfonts}
\usepackage{rotating}
\usepackage{cancel}
\usepackage{mathtools}
\usepackage{bbm}
\usepackage{dsfont}
\usepackage{bbold}
\usepackage{multirow}
\usepackage{ulem}
\usepackage{physics}
\usepackage{orcidlink}
\usepackage{colortbl}

\def\pslash{p\!\!\!\slash }
\def\p_1slash{p1\!\!\!\slash }
\def\p_2slash{p2\!\!\!\slash }
\def\qslash{q\!\!\!\slash }
\def\xslash{x\!\!\!\slash }
\def\eslash{\varepsilon\!\!\!\slash }

\def\vel{\left|}
\def\ver{\right|}

\begin{document}

\title{Shedding light on the nature of the $P_{cs}(4459)$ pentaquark state}

\author{Ula\c{s}~\"{O}zdem\orcidlink{0000-0002-1907-2894}}%
\email[]{ulasozdem@aydin.edu.tr }
\affiliation{ Health Services Vocational School of Higher Education, Istanbul Aydin University, Sefakoy-Kucukcekmece, 34295 Istanbul, T\"{u}rkiye}

\begin{abstract}
To shed light on the properties of states whose nature, internal structure, and spin-parity quantum numbers are not fully elucidated, we systematically study their electromagnetic properties.  In light of this concept, we present a comprehensive analysis of the magnetic dipole moment of the $P_{cs}(4459)$ pentaquark within the context of QCD light-cone sum rules, utilizing three distinct interpolating currents in the form of diquark-diquark-antiquark configurations that are likely to couple this pentaquark with $J^P =\frac{3}{2}^-$ quantum numbers.  The numerical analysis yielded the following results:  $\mu_{{J_\mu^1}}= -0.60 \pm 0.15~\mu_N$, $\mu_{{J_\mu^2}}=1.60 \pm 0.30~\mu_N$ , and $\mu_{{J_\mu^3}}= 0.99 \pm 0.20~\mu_N$. 
The numerical results obtained have led to the conclusion that the magnetic dipole moments of the $P_{cs}(4459)$ state are capable of projecting its inner structure. As is seen, the different diquark-diquark-antiquark configurations of the $P_{cs}(4459)$ pentaquark state contain important information about its internal structure. Thus, this study will provide prominent data to investigate the inner structure of the $P_{cs}(4459)$ pentaquark state. We compared our results with other theoretical predictions that could be a useful complementary tool for interpreting the nature of the $P_{cs}(4459)$ state. 
A thorough  examination reveals that the results obtained by employing disparate theoretical approaches and different internal structure models are not consistent with each other.  
It is recommended that further studies be conducted using alternative non-perturbative techniques to gain a more comprehensive understanding of the observed results.
\end{abstract}

\maketitle

\section{Introduction}\label{motivation}

While the existence of states with a structure distinct from that of standard baryons and mesons has been postulated for several decades, the first experimental confirmation of such states was achieved in 2003 by the Belle Collaboration~\cite{Belle:2003nnu}.  This state, which has been identified through experimental means and is known as X(3872), has been classified as the first exotic tetraquark state due to its anomalous properties, which are inconsistent with those expected of a conventional meson state. Since that time, the number of exotic multi-quark states that have been observed through experimental means has increased, and the diversity of these states has also grown. As a result of these experimental findings, the study of the properties of these unconventional multi-quark states has become a prominent and intriguing area of research within the field of hadron physics. The investigation of the characteristics of these hadrons may offer significant insights into the underlying mechanisms of strong interactions at low energies. Several theoretical models have been put forth to shed light on these experimental observations, offering a variety of insights into the characteristics of these states. These include the compact multiquark configurations, molecular bound states, kinematic effects, and so on ~\cite{Esposito:2014rxa, Esposito:2016noz, Olsen:2017bmm, Lebed:2016hpi, Nielsen:2009uh, Brambilla:2019esw, Agaev:2020zad, Chen:2016qju, Ali:2017jda, Guo:2017jvc, Liu:2019zoy, Yang:2020atz, Dong:2021juy, Dong:2021bvy, Chen:2022asf, Meng:2022ozq}. 
 
In 2015, the LHCb Collaboration made a remarkable breakthrough in the study of exotic multi-quark states with the discovery of two pentaquarks, designated $P_{c }(4380)$ and $P_{c }(4450)$, in the $J/\psi p$ invariant mass spectrum of the process $ \Lambda^0_b \rightarrow J/\psi pK^-$. Subsequently, the LHCb Collaboration updated their analysis of the $J/\psi p$ invariant mass distribution of the decay $ \Lambda_b \rightarrow J/\psi pK $, utilizing the data collected in Run I and Run II. In the updated analysis, a new pentaquark state, $P_c(4312)$, was identified, while the $P_c(4450)$ split into two structures, which were designated as $P_c(4440)$ and $P_c(4457)$, respectively. It should be noted that the pentaquark $P_c(4380)$ reported in the previous analysis remains unresolved, neither confirmed nor refuted, in the subsequent analysis. Since these pentaquarks are observed in the $J/\psi p$ invariant mass distribution, it is presumed that their quark content is $uudc\bar c$. Building upon these observations, the LHCb Collaboration advanced their understanding of hidden-charm pentaquark states with strangeness. In 2020, the LHCb Collaboration reported evidence of a pentaquark candidate, designated as $P_{cs}(4459)$, in the $J/\psi \Lambda$ invariant mass spectrum in the decay $\Xi^-_b  \rightarrow J/\psi \Lambda K^- $.   In 2022, a pentaquark candidate, designated $P_{cs}(4338)$, was also reported by the LHCb in the $J/\psi\Lambda$ invariant mass spectrum.
Given that $P_{cs}(4459)$ and $P_{cs}(4338)$ states have been observed in the invariant mass distribution $J/\psi\Lambda$, the assumptions have been made that their quark content is $udsc\bar c$.
 The masses and  widths for the LHCb pentaquark states have been reported as follows \cite{LHCb:2015yax, LHCb:2019kea, LHCb:2020jpq, LHCb:2022ogu}:
\begin{eqnarray}
{P_c(4380)}:~~ \rm{M}&=&4380 \pm 8 \pm 29~\mbox{MeV} ~~~~~~~~~~ ~~ 
\Gamma=205 \pm 18 \pm 86~\mbox{MeV}, \nonumber\\
 P_c(4440):~~ \rm{M}&=&4440.3 \pm 1.3 ^{+ 4.1}_{-4.7}~\mbox{MeV} ~~~~~~~~~  \Gamma= 20.6 \pm 4.9^{+8.7}_{-10.1}~\mbox{MeV},\nonumber\\
P_c(4457):~~ \rm{M}&=&4457.3 \pm 0.6 ^{+ 4.1}_{-1.7}~\mbox{MeV} ~~~~~~~~~  \Gamma= 6.4 \pm 2.0^{+5.7}_{-1.9}~\mbox{MeV},\nonumber\\
 P_c(4312):~~ \rm{M}&=&4311.9 \pm 0.7^{ +6.8}_{-0.6}~\mbox{MeV} ~~~~~~~~~  \Gamma=9.8 \pm 2.7 ^{ +3.7}_{-4.5}~\mbox{MeV},\nonumber\\
 P_{cs}(4459):~~ \rm{M}&=&4458.8 \pm 2.7 ^{+4.7}_{-1.1}~\mbox{MeV} ~ ~~~~~~~~  \Gamma =17.3 \pm 6.5^{+8.0}_{-5.7}~\mbox{MeV},\nonumber\\
 P_{cs}(4338):~~ \rm{M} &=&4338.2 \pm 0.7 \pm 0.4~\mbox{MeV} ~~~~~~~
\Gamma=7.0 \pm 1.2 \pm 1.3~\mbox{MeV}.\nonumber
\end{eqnarray}

Very recently, Belle Collaboration  find evidence of the $P_{cs}(4459)$ state with a significance of 3.3$\sigma$,    including statistical and systematic uncertainties. They  measure the mass and width of the $P_{cs}(4459)$ to be $(4471.7 \pm 4.8 \pm 0.6)~$ MeV and $(21.9 \pm 13.1 \pm 2.7)~$ MeV, respectively \cite{Belle:2025pey}.
The spin-parity quantum numbers of these states have yet to be elucidated.  Despite the considerable efforts made since the initial experimental observations to elucidate not only the quantum numbers but also the internal structure of these states, their true nature remains an open question that requires further investigation. 
It is therefore essential to study various properties such as their decay constants, decay branching ratios, and transition form factors, as these properties provide unique and detailed insights into their internal structure. Magnetic moments are a simple test of nonperturbative models and are valuable for probing the internal structure of exotic states through their electromagnetic interactions.
 
In this study, we present a comprehensive analysis of the magnetic dipole moment of the $P_{cs}(4459)$ pentaquark within the context of QCD light-cone sum rules, utilizing three distinct interpolating currents in the form of diquark-diquark-antiquark configurations that are likely to couple this pentaquark with $J^P =\frac{3}{2}^-$ quantum numbers. As is well known, magnetic dipole moments are physical quantities directly correlated with the internal structure of the hadron under investigation.  They can provide insights into the internal structure of the hadron and the low-energy region of the QCD. Furthermore, magnetic dipole moments represent an effective tool for investigating the dynamics of quarks and gluons within a hadron. This is because it represents the leading-order response of a bound state to an external magnetic field. The existing literature has been the subject of increasing interest as a result of the invaluable insights it offers into studies that employ a variety of models and techniques to investigate the magnetic dipole moments of hidden-charm/bottom pentaquarks~\cite{Wang:2016dzu, Ozdem:2018qeh, Ortiz-Pacheco:2018ccl, Xu:2020flp, Ozdem:2021btf, Ozdem:2021ugy, Li:2021ryu, Ozdem:2023htj, Wang:2023iox, Ozdem:2022kei, Gao:2021hmv, Guo:2023fih, Ozdem:2022iqk, Wang:2022nqs, Wang:2022tib, Ozdem:2024jty, Li:2024wxr, Li:2024jlq, Ozdem:2024yel, Ozdem:2024rqx, Mutuk:2024ltc, Mutuk:2024jxf, Ozdem:2024usw}. Although the short lifetime of the $P_{cs}$ state currently presents a challenge for measuring the magnetic dipole moment, the accumulation of more data from experiments in the future may enable this to be achieved.   The $\Delta^+(1232)$ baryon has also a very short lifetime, however, its magnetic dipole moment was achieved through $\gamma N $ $ \rightarrow $ $ \Delta $ $\rightarrow $ $ \Delta \gamma $ $ \rightarrow$ $ \pi N \gamma $ process~ \cite{Pascalutsa:2004je, Pascalutsa:2005vq, Pascalutsa:2007wb}. A comparable $\gamma^{(*)}\Lambda $ $ \rightarrow $ $P_{cs} $ $\rightarrow P_{cs} \gamma$ $ \rightarrow $ $ J/\psi \Lambda \gamma$ process may be employed to derive the magnetic dipole moment of the $P_{cs}$ pentaquark.  In addition, the magnetic dipole moments of doubly charmed baryons have been calculated using lattice QCD~\cite{Can:2013zpa, Can:2013tna}.  It may be feasible to extend these analyses to encompass tetra- and -pentaquark states in the near future.

We organize this paper in the following manner:  The following section will present a methodology for identifying potential interpolating currents for the pentaquark $P_{cs}$, which will then be employed to derive sum rules for the desired physical parameter. The numerical results and the relevant discussions are presented in Sec.~\ref{numerical}. Finally, this work ends with the summary in Sec.~\ref{summary}.

\begin{widetext}
 
\section{Theoretical Frame}\label{formalism}

 The QCD light-cone sum rule is a well-established and effective method for determining the observable characteristics of hadrons, including their form factors, strong and weak decays, and radiative decays. Furthermore, the technique provides valuable insights into the internal structure of hadrons~\cite{Chernyak:1990ag, Braun:1988qv, Balitsky:1989ry}. The method computes the correlation function, which constitutes a fundamental component of the method, through two distinct ways: the QCD  and the hadronic representations.  At the hadron level, hadronic parameters, such as residues and form factors, are employed, whereas at the QCD  level, parameters associated with QCD, such as the quark condensate, distribution amplitudes of the corresponding particle, and so forth, are utilized. Subsequently, to eliminate the contributions of the unwanted effects such as the continuum and higher states, the double Borel transformation is performed on both representations regarding the parameters $p^2$ and $(p+q)^2$. Thereafter, the quark-hadron duality approach is applied. As a result of these procedures, sum rules for the corresponding parameters, which in our case is the magnetic dipole moment, are obtained.

 Having provided a brief overview of the method, we may now proceed to analyze the physical parameter in question using this approach. As previously stated, to accomplish this, it is first necessary to define the relevant correlation function. The correlation function to be employed in the analysis of the magnetic dipole moment of the $P_{cs}$ state is provided by the following formula \cite{Ball:2002ps,Aliev:2008sk,Azizi:2009egn,Aliev:2014foa,Ozdem:2024brk}:
 
\begin{eqnarray} \label{edmn01}
\Pi_{\mu \nu}(p,q)&=&i\int d^4x e^{ip \cdot x} \langle0|T\left\{J_{\mu}(x)\bar{J}_{\nu}(0)\right\}|0\rangle _F \, ,
\end{eqnarray}
where  the $J_{\mu}(x)$ stands for interpolating current of the $P_{cs}$ state and   $F$ means the external electromagnetic field. 

One effective way for characterizing hadron properties is to assign a specific structure, subsequently analyzing the resulting hadron properties based on that structure. The results of these analyses can provide significant insights into the nature, quantum numbers, and internal structure of the hadron in question.  It is of paramount importance to select suitable interpolating currents, composed of quark fields with an identical quark content and quantum numbers as those of the $P_{cs}$ state, to achieve this. Based on this, all the possible interpolating currents that would couple the relevant hadron are written down. Given our focus on interpolating currents within the diquark-diquark-antiquark configuration, the construction of suitable diquark structures becomes crucial. The diquarks $q^{T}_a C\Gamma q^{\prime}_b$ have  five  structures  in Dirac spinor space, where $C\Gamma=C\gamma_5$, $C$, $C\gamma_\mu \gamma_5$,  $C\gamma_\mu $ and $C\sigma_{\mu\nu}$ for the scalar, pseudoscalar, vector, axialvector  and  tensor diquarks, respectively, and the $a$ and $b$ are color indexes. The attractive interactions of one-gluon exchange  favor  formation of
the diquarks in  color antitriplet $\overline{3}_{ c}$, flavor
antitriplet $\overline{3}_{ f}$ and spin singlet $1_s$, while the favored configurations are the scalar ($C\gamma_5$) and axialvector ($C\gamma_\mu$) diquark states. The QCD sum rules indicate that the scalar   and axial-vector  diquark configurations are the most favorable for investigating the characteristics of hadrons~\cite {Wang:2010sh, Kleiv:2013dta}. In light of the considerations mentioned earlier, we have introduced to utilize the axialvector-diquark-scalar-diquark-antiquark and the axialvector-diquark-axialvector-diquark-antiquark types of interpolating currents in the course of this investigation. The relevant expressions are presented as follows:
\begin{eqnarray}
J_{\mu}^{1}(x)&=&\frac{\mathcal A }{\sqrt{3}} \bigg\{ \big[ {u}^T_d(x) C \gamma_\mu {d}_e(x) \big] \big[ {s}^T_f(x) C \gamma_5 c_g(x)\big]  C  \bar{c}^{T}_{c}(x) +
\big[ {u}^T_d(x) C \gamma_\mu {s}_e(x) \big] \big[ {d}^T_f(x) C \gamma_5 c_g(x)\big]  C  \bar{c}^{T}_{c}(x) \nonumber\\
&& +\big[ {d}^T_d(x) C \gamma_\mu {s}_e(x) \big] \big[ {u}^T_f(x) C \gamma_5 c_g(x)\big]  C  \bar{c}^{T}_{c}(x)\bigg\} \, , \\
J_{\mu}^{2}(x)&=& \frac{\mathcal A }{\sqrt{3}} \bigg\{ \big[ {u}^T_d(x) C \gamma_\mu {d}_e(x) \big] \big[ {s}^T_f(x) C \gamma_\alpha c_g(x)\big]  \gamma_5 \gamma^\alpha C  \bar{c}^{T}_{c}(x) +
\big[ {u}^T_d(x) C \gamma_\mu {s}_e(x) \big] \big[ {d}^T_f(x) C \gamma_\alpha c_g(x)\big]  \gamma_5 \gamma^\alpha C \bar{c}^{T}_{c}(x) \nonumber\\
&& +\big[ {d}^T_d(x) C \gamma_\mu {s}_e(x) \big] \big[ {u}^T_f(x) C \gamma_\alpha c_g(x)\big]  \gamma_5 \gamma^\alpha C  \bar{c}^{T}_{c}(x)\bigg\}  \, , \\
J_{\mu}^{3}(x)&=& \frac{\mathcal A }{\sqrt{3}} \bigg\{ \big[ {u}^T_d(x) C \gamma_\alpha {d}_e(x) \big] \big[ {s}^T_f(x) C \gamma_\mu c_g(x)\big]  \gamma_5 \gamma^\alpha C  \bar{c}^{T}_{c}(x) + \big[ {u}^T_d(x) C \gamma_\alpha {s}_e(x) \big] \big[ {d}^T_f(x) C \gamma_\mu c_g(x)\big]  \gamma_5 \gamma^\alpha C \bar{c}^{T}_{c}(x) \nonumber\\ && +\big[ {d}^T_d(x) C \gamma_\alpha {s}_e(x) \big] \big[ {u}^T_f(x) C \gamma_\mu c_g(x)\big]  \gamma_5 \gamma^\alpha C  \bar{c}^{T}_{c}(x)\bigg\} \, , 
 \end{eqnarray}
where $ \mathcal A = \varepsilon_{abc}\varepsilon_{ade} \varepsilon_{bfg}$ with   
$a$, $b$, $c$, $d$, $e$, $f$ and $g$ being color indices; and the $C$ is the charge conjugation operator. As can be observed, the interpolating currents presented above possess the same quantum number and quark content. It may therefore be anticipated that they will couple to the same pentaquark state in principle. 
At the hadronic level, a significant distinction emerges between molecular states and compact pentaquark states. Molecular states are widely presumed to be bound states of two hadrons, formed through the exchange of a color-singlet meson and a baryon. In contrast, compact pentaquark states are typically bound by the strong force at the quark-gluon level within the framework of QCD. However, within the framework of the so-called "QCD sum rule approach," the only difference lies in the interpolating current used in the study of the molecular and compact pentaquark states.   In principle, if we examine all the possible molecular-type currents and all the possible compact pentaquark-type currents, we can rigorously show that these two sets of interpolating currents are equivalent by using the so-called "Fierz rearrangement." However, a notable distinction exists between a single molecular-type current and a single compact pentaquark-type current. For instance, every pentaquark-type current is a linear combination of multiple independent molecular-type currents. In this respect, one well-known example is the light scalar-isoscalar sigma meson, where the tetraquark-type current or its combination/mixing yields a more favorable result compared to the conventional pion-pion molecular current. Distinguishing between the compact pentaquark and molecular structures is a potential outcome of our present systematical investigation, motivated by the necessity to understand the nature of this state.

At the hadron level, we insert a complete set of intermediate hadronic states with the same quantum numbers as the interpolating currents $J_\mu^1(x)$, $J_\mu^2(x)$ and $J_\mu^3(x)$ into the correlation function in Eq. (\ref{edmn01}) under the principles of quark-hadron duality. This allows us to obtain the hadronic spectral representation. Subsequently, the ground state contributions of the $P_{cs}$ state are isolated. The result of this procedure is as follows.

\begin{eqnarray}\label{edmn02}
\Pi^{Had}_{\mu\nu}(p,q)&=&\frac{\langle0\mid J_{\mu}(x)\mid
P_{cs}(p_2,s)\rangle}{[p_2^{2}-m_{P_{cs}}^{2}]}\langle P_{cs}(p_2,s)\mid
P_{cs}(p_1,s)\rangle_F\frac{\langle P_{cs}(p_1,s)\mid
\bar{J}_{\nu}(0)\mid 0\rangle}{[p_1^{2}-m_{P_{cs}}^{2}]} \nonumber\\
&&+ \mbox{ higher states and continuum},
\end{eqnarray}
where $p_1 = p+q$, $p_2=p$, and $q$ is the momentum of the photon. 
As illustrated in Eq.~(\ref{edmn02}), explicit expressions of $\langle0\mid J_{\mu}(x)\mid P_{cs}(p_2,s)\rangle$, $\langle {P_{cs}}(p_1,s)\mid
\bar{J}_{\nu}^{P_{cs}}(0)\mid 0\rangle$, and $\langle P_{cs}(p_2,s)\mid P_{cs}(p_1,s)\rangle_F$ matrix elements are required. These are expressed as follows:

\begin{eqnarray}
\label{lambdabey}
\langle0\mid J_{\mu}(x)\mid P_{cs}(p_2,s)\rangle &=&\lambda_{P_{cs}}u_{\mu}(p_2,s),\\
\langle {P_{cs}}(p_1,s)\mid
\bar{J}_{\nu}(0)\mid 0\rangle &=& \lambda_{{P_{cs}}}\bar u_{\nu}(p_1,s),\\
\langle P_{cs}(p_2,s)\mid P_{cs}(p_1,s)\rangle_F &=&-e \,\bar
u_{\mu}(p_2,s)\left\{F_{1}(q^2)g_{\mu\nu}\eslash-
\frac{1}{2m_{P_{cs}}}\left
[F_{2}(q^2)g_{\mu\nu} \eslash\qslash+F_{4}(q^2)\frac{q_{\mu}q_{\nu} \eslash\qslash}{(2m_{P_{cs}})^2}\right]
\right.\nonumber\\&+&\left.
\frac{F_{3}(q^2)}{(2m_{P_{cs}})^2}q_{\mu}q_{\nu}\eslash\right\} u_{\nu}(p_1,s),\label{matelpar}
\end{eqnarray}
where $\lambda_{P_{cs}}$  is pole residue of the $P_{cs}$ state, which is a constant that parameterizes the coupling strength of the hadron to the current; $u_{\mu}(p_2,s)$ and $ \bar u_{\nu}(p_1,s)$ are the Rarita-Schwinger spinor vectors; $\varepsilon$ is the photon's polarization vector, and $F_i (q^2)$ are transition form factors.  By employing the Eqs. (\ref{edmn02})-(\ref{matelpar}), the correlation function in terms of hadronic parameters is derived.   Once these steps have been completed, the final version of the function is as follows: 
\begin{align}\label{fizson}
 \Pi^{Had}_{\mu\nu}(p,q)&=-\frac{\lambda_{{P_{cs}}}^{2}}{[p_1^{2}-m_{_{P_{cs}}}^{2}][p_2^{2}-m_{_{P_{cs}}}^{2}]}
 \big(\pslash_1+m_{P_{cs}}\big)
 \bigg[g_{\mu\nu}
-\frac{1}{3}\gamma_{\mu}\gamma_{\nu}-\frac{2\,p_{1\mu}p_{1\nu}}
{3\,m^{2}_{P_{cs}}}+\frac{p_{1\mu}\gamma_{\nu}-p_{1\nu}\gamma_{\mu}}{3\,m_{P_{cs}}}\bigg] \bigg\{F_{1}(q^2)g_{\mu\nu}\eslash  \nonumber\\
& -
\frac{1}{2m_{P_{cs}}}
\Big[F_{2}(q^2)g_{\mu\nu}\eslash\qslash +F_{4}(q^2) \frac{q_{\mu}q_{\nu}\eslash\qslash}{(2m_{P_{cs}})^2}\Big]+\frac{F_{3}(q^2)}{(2m_{P_{cs}})^2}
 q_{\mu}q_{\nu}\eslash\bigg\}
 \big(\pslash_2+m_{P_{cs}}\big)
 \bigg[g_{\mu\nu}-\frac{1}{3}\gamma_{\mu}\gamma_{\nu}-\frac{2\,p_{2\mu}p_{2\nu}}
{3\,m^{2}_{P_{cs}}}\nonumber\\
&+\frac{p_{2\mu}\gamma_{\nu}-p_{2\nu}\gamma_{\mu}}{3\,m_{P_{cs}}}\bigg].
\end{align}

In our calculation, we also performed summation over spins of the Rarita-Schwinger spin vector,  
\begin{align}\label{raritabela}
\sum_{s}u_{\mu}(p,s)\bar u_{\nu}(p,s)=-\big(\pslash+m_{P_{cs}}\big)\Big[g_{\mu\nu}
-\frac{1}{3}\gamma_{\mu}\gamma_{\nu}-\frac{2\,p_{\mu}p_{\nu}}
{3\,m^{2}_{P_{cs}}}+\frac{p_{\mu}\gamma_{\nu}-p_{\nu}\gamma_{\mu}}{3\,m_{P_{cs}}}\Big].
\end{align}

In principle, the equations mentioned above could be employed to derive the hadronic level of the analysis. However, at this juncture, two issues must be addressed, as they have the potential to impact the reliability of the calculations.  Firstly, it should be noted that not all Lorentz structures appearing in Eq. (\ref{fizson}) are independent. Secondly, the correlation function contains contributions not only from spin-3/2 but also from spin-1/2 pentaquark states that must be eliminated. The matrix element of the vacuum and the interpolating current between the spin-1/2 hadrons can be defined as follows:
\begin{equation}\label{spin12}
\langle0\mid J_{\mu}(0)\mid B(p,s=1/2)\rangle=(A  p_{\mu}+B\gamma_{\mu})u(p,s=1/2).
\end{equation}
As illustrated by this equation, the unwanted effects concerning spin-1/2 hadrons are proportional to both $p_\mu$ and $\gamma_\mu$.  
To eliminate the unwanted pollution originating from the spin-1/2 pentaquarks and to obtain only independent structures within the correlation function, we implement the ordering of Dirac matrices as follows: $\gamma_{\mu}\pslash\eslash\qslash\gamma_{\nu}$.  Then, to ensure that we have removed these unwanted contributions from the analysis, we need to remove terms with $\gamma_\mu$ at the beginning and $\gamma_\nu$ at the end, or that is proportional to $p_{2\mu}$ or $p_{1\nu}$~\cite{Belyaev:1982cd}. The result of the analysis at the hadron level, after performing the above procedures, is in the following form:

\begin{eqnarray}
\label{final phenpart}
\Pi^{Had}_{\mu\nu}(p,q)&=&\frac{\lambda_{_{P_{cs}}}^{2}}{[(p+q)^{2}-m_{_{P_{cs}}}^{2}][p^{2}-m_{_{P_{cs}}}^{2}]}
\Bigg[  g_{\mu\nu}\pslash\eslash\qslash \,F_{1}(q^2) 
-m_{P_{cs}}g_{\mu\nu}\eslash\qslash\,F_{2}(q^2)+
\frac{F_{3}(q^2)}{2m_{P_{cs}}}q_{\mu}q_{\nu}\eslash\qslash\, \nonumber\\
&+&
\frac{F_{4}(q^2)}{4m_{P_{cs}}^3}(\varepsilon.p)q_{\mu}q_{\nu}\pslash\qslash \,+
\mathrm{other~independent~structures} \Bigg].
\end{eqnarray}

It may be more useful to write the form factors given in Eq.~(\ref{final phenpart}) in terms of the magnetic form factor, designated as $G_{M}(q^2)$, which is more understandable and easier to study experimentally. Therefore, we need the form factor $G_{M}(q^2)$ written in terms of the form factors $F_{i}(q^2)$, which is prescribed as follows~\cite{Weber:1978dh,Nozawa:1990gt,Pascalutsa:2006up,Ramalho:2009vc}: 

\begin{eqnarray}
G_{M}(q^2) &=& \left[ F_1(q^2) + F_2(q^2)\right] ( 1+ \frac{4}{5}
\tau ) -\frac{2}{5} \left[ F_3(q^2)  +
F_4(q^2)\right] \tau \left( 1 + \tau \right), 
\end{eqnarray}
  where $\tau
= -\frac{q^2}{4m^2_{P_{cs}}}$.  In the static limit, that is to say, when $q^2=0$, the magnetic form factor associated with the $F_i(0)$ form factors is described by the following formula:

\begin{eqnarray}\label{mqo1}
G_{M}(0)&=&F_{1}(0)+F_{2}(0).
\end{eqnarray}
 
Since our study focuses on the magnetic dipole moment analysis, it is required to formulate the magnetic dipole moment regarding the aforementioned form factors. However, this can only be done in the context of the static limit. 
The magnetic dipole moment, designated as ($\mu_{P_{cs}}$), can be extracted as follows with the help of the equation given above as: 

 \begin{eqnarray}\label{mqo2}
\mu_{P_{cs}}&=&\frac{e}{2m_{P_{cs}}}G_{M}(0).
\end{eqnarray}

It should be noted that following the specified quark content, there exists the possibility of having more than one $P_{cs}$ state that is capable of coupling to a given interpolating current. If the masses of the $P_{cs}$ states exhibit a significant disparity, the equations presented in Eqs. (\ref{edmn02})-(\ref{mqo2}) can be applied to the lowest mass $P_{cs}$ state. In such a case, the magnetic dipole moments derived through the use of distinct interpolating currents should yield the same value. However, If the masses of these $P_{cs}$ states are similar to one another, it means that they are close to degenerate states. In that scenario, Eq. (\ref{mqo2}) provides a weighted average where the weighing computed by the residues squares, of the magnetic dipole moments of these $P_{cs}$ states. It is important to note that the magnetic dipole moments achieved through the utilization of varying interpolating currents may exhibit discrepancies if the magnetic dipole moments of these nearly degenerate $P_{cs}$ states exhibit notable differences. This outcome can be attributed to the fact that the magnetic dipole moments are derived through the utilization of distinct weight factors for varying interpolating currents.

At the quark-gluon level of the sum rules, the correlation function is derived from the quark propagators and analyzed with some accuracy through the operator product expansion based on photon distribution amplitudes. Once the pertinent quark fields have been contracted by Wick's theorem, the resulting correlation function is an expression for the quark propagators, as detailed in the following expression. 
For the sake of simplicity, the result for the $J_\mu^1$ interpolating current is presented here. This result is given as follows: 

\begin{eqnarray}
\label{QCD1}
\Pi^{QCD,\,J_\mu^1}_{\mu\nu}(p,q)&=& \frac{i}{3}\,\varepsilon^{abc}\varepsilon^{a^{\prime}b^{\prime}c^{\prime}}\varepsilon^{ade}
\varepsilon^{a^{\prime}d^{\prime}e^{\prime}}\varepsilon^{bfg}
\varepsilon^{b^{\prime}f^{\prime}g^{\prime}} \int d^4x e^{ip\cdot x}\langle 0|
\Big\{
 \nonumber\\
&&- \mbox{Tr}\Big[  \gamma_\mu S_d^{ee^\prime}(x) \gamma_\nu  C S_u^{dd^\prime \mathrm{T}}(x) C\Big]
 \mbox{Tr}\Big[ \gamma_5 S_c^{gg^\prime}(x) \gamma_5 C  S_s^{ff^\prime \mathrm{T}}(x)C \Big] 
 \nonumber\\
&&
- \mbox{Tr}\Big[  \gamma_\mu S_s^{ee^\prime}(x) \gamma_\nu C  S_u^{dd^\prime \mathrm{T}}(x) C\Big]
 \mbox{Tr}\Big[ \gamma_5 S_c^{gg^\prime}(x) \gamma_5 C  S_d^{ff^\prime \mathrm{T}}(x)C  \Big] 
 \nonumber\\
&&
- \mbox{Tr}\Big[  \gamma_\mu S_s^{ee^\prime}(x) \gamma_\nu C  S_d^{dd^\prime \mathrm{T}}(x)C\Big]
 \mbox{Tr}\Big[ \gamma_5 S_c^{gg^\prime}(x) \gamma_5 C  S_u^{ff^\prime \mathrm{T}}(x)C \Big] 
 \nonumber\\
&&
+   \mbox{Tr} \Big[  \gamma_5 S_c^{gg^\prime}(x) 
\gamma_5 C S_d^{ef^\prime \mathrm{T}}(x)  C \gamma_\mu S_u^{dd^\prime}(x) \gamma_\nu C  S_s^{fe^\prime \mathrm{T}}(x)C\Big]\nonumber\\
&&
-   \mbox{Tr} \Big[  \gamma_5 S_c^{gg^\prime}(x) 
\gamma_5 C S_u^{df^\prime \mathrm{T}}(x)  C \gamma_\mu S_d^{ed^\prime}(x) \gamma_\nu C  S_s^{fe^\prime \mathrm{T}}(x)C\Big]
\nonumber\\
&&
+  \mbox{Tr} \Big[  \gamma_5 S_c^{gg^\prime}(x) 
\gamma_5 C S_s^{ef^\prime \mathrm{T}}(x)  C \gamma_\mu S_u^{dd^\prime}(x) \gamma_\nu C  S_d^{fe^\prime \mathrm{T}}(x)C\Big]
\nonumber\\
&&
 + \mbox{Tr} \Big[  \gamma_5 S_c^{gg^\prime}(x) 
\gamma_5 C S_u^{df^\prime \mathrm{T}}(x)  C \gamma_\mu S_s^{ee^\prime}(x) \gamma_\nu C  S_d^{fd^\prime \mathrm{T}}(x)C\Big]\nonumber\\
&&
 - \mbox{Tr} \Big[  \gamma_5 S_c^{gg^\prime}(x) 
\gamma_5 C S_s^{ef^\prime \mathrm{T}}(x)  C \gamma_\mu S_d^{ed^\prime}(x) \gamma_\nu C  S_u^{fd^\prime \mathrm{T}}(x)C\Big]\nonumber\\
&&
 + \mbox{Tr} \Big[ \gamma_5 S_c^{gg^\prime}(x) 
\gamma_5 C S_d^{df^\prime \mathrm{T}}(x)  C \gamma_\mu S_s^{ee^\prime}(x) \gamma_\nu C  S_u^{fd^\prime \mathrm{T}}(x) C\Big]
\Big \} \Big(C S_c^{c^{\prime}c \mathrm{T}} (-x) C \Big)
|0 \rangle_F .
\end{eqnarray}

 The $S_{q}(x)$ and $S_{c}(x)$ in Eq. (\ref{QCD1})is the corresponding light and heavy  propagators, which are provided  as~\cite{Balitsky:1987bk, Belyaev:1985wza}:
\begin{align}
\label{edmn13}
S_{q}(x)&= S_q^{free}(x) - \frac{\langle \bar qq \rangle }{12} \Big(1-i\frac{m_{q} \xslash}{4}   \Big)- \frac{ \langle \bar qq \rangle }{192}
m_0^2 x^2  \Big(1-i\frac{m_{q} \xslash}{6}   \Big)
+\frac {i g_s }{16 \pi^2 x^2} \int_0^1 dv \, G^{\mu \nu} (vx)
\bigg[\bar v \rlap/{x} 
\sigma_{\mu \nu} + v \sigma_{\mu \nu} \rlap/{x}
 \bigg],\\
S_{Q}(x)&=S_Q^{free}(x)
-\frac{m_{Q}\,g_{s} }{16\pi ^{2}}  \int_0^1 dv \,G^{\mu \nu}(vx)\bigg[  
    \frac{K_{1}\big( m_{Q}\sqrt{-x^{2}}\big) }{\sqrt{-x^{2}}}(\sigma _{\mu \nu }{\xslash}
+{\xslash}\sigma _{\mu \nu })
 +2\sigma_{\mu \nu }K_{0}\big( m_{Q}\sqrt{-x^{2}}\big)\bigg],
 \label{edmn14}
\end{align}%
with  
\begin{align}
 S_q^{free}(x)&=\frac{1}{2 \pi x^2}\Big(i \frac{\xslash}{x^2}- \frac{m_q}{2}\Big),\\
 \nonumber\\
 S_c^{free}(x)&=\frac{m_{c}^{2}}{4 \pi^{2}} \bigg[ \frac{K_{1}\big(m_{c}\sqrt{-x^{2}}\big) }{\sqrt{-x^{2}}}
+i\frac{{\xslash}~K_{2}\big( m_{c}\sqrt{-x^{2}}\big)}
{(\sqrt{-x^{2}})^{2}}\bigg].
\end{align}
where $G^{\mu\nu}(x)$ is the gluon field strength tensor,  $v$ is the line variable, and $K_n(m_Q\sqrt{-x^2})$ are the Bessel functions. Here, we use the following form  of the  Bessel function,     
\begin{equation}\label{b2}
K_n(m_Q\sqrt{-x^2})=\frac{\Gamma(n+ 1/2)~2^n}{m_Q^n \,\sqrt{\pi}}\int_0^\infty \cos(m_Qt)\frac{(\sqrt{-x^2})^n}{(t^2-x^2)^{n+1/2}}~dt.
\end{equation}

In the representation of the correlation function at the quark-gluon level, two distinct contributions emerge, designated as perturbative and non-perturbative.  It is essential to consider the contributions from these two distinct regions for the analysis to be both reliable and consistent. The perturbative contributions are regarding the short-distance interaction of the photon with both light and heavy quarks, whereas the non-perturbative contributions concern the long-distance interaction of the photon with quark fields.

To incorporate perturbative contributions into the calculations, one of the light or heavy quark propagators that are in interaction with the photon should be substituted according to the following replacement:
\begin{align}
\label{free}
S_{q(c)}^{free}(x) \rightarrow \int d^4z\, S_{q(c)}^{free} (x-z)\,\rlap/{\!A}(z)\, S_{q(c)}^{free} (z)\,.
\end{align}

To incorporate non-perturbative contributions into the calculations, it is necessary to substitute one of the light quark propagators interacting with the photon at a long distance, under the following substitution:
\begin{align}
\label{edmn15}
S_{q, \alpha\beta}^{ab}(x) \rightarrow -\frac{1}{4} \big[\bar{q}^a(x) \Gamma_i q^b(0)\big]\big(\Gamma_i\big)_{\alpha\beta},
\end{align}
where   $\Gamma_i = \mathrm{1}, \gamma_5, \gamma_\mu, i\gamma_5 \gamma_\mu, \sigma_{\mu\nu}/2$. 

Upon analysis of both the perturbative and non-perturbative contributions, it is determined that one of the quarks interacts with the photon. The remaining four propagators are incorporated into the non-perturbative analysis as full quark propagators, whereas the perturbative analysis employs only the free part.  When a photon interacts non-perturbatively with light-quark fields, expressions such as $\langle \gamma(q)\vel \bar{q}(x) \Gamma_i G_{\mu\nu}q(0) \ver 0\rangle$  and $\langle \gamma(q)\vel \bar{q}(x) \Gamma_i q(0) \ver 0\rangle$ emerge and are expressed in terms of photon distribution amplitudes (DAs)~\cite{Ball:2002ps}.  This is because this part of the calculation is a technical and standardized process, we refrain from providing further detail here; however, interested readers may refer to the Refs.~\cite{Ozdem:2022vip, Ozdem:2022eds} for more comprehensive information about these procedures. It is crucial to emphasize that the photon DAs employed in this analysis encompass exclusively contributions from light quarks. Nevertheless, it is theoretically possible for a photon to be emitted from charm quarks over a long distance. However, such effects are significantly suppressed due to their large mass. This type of contribution is disregarded in our analysis. As outlined in Eq. (\ref{free}), only the short-distance photon emission from heavy quarks is incorporated. Consequently, DAs incorporating charm quarks are not included in our evaluations.   With the help of Eqs.~(\ref{free}) and (\ref{edmn15}), both perturbative and non-perturbative contributions have been included in the analysis according to the standard prescription of the method. The correlation function in terms of quark-gluon parameters is derived through the aforementioned manipulations, followed by the application of the Fourier transform to transfer the expressions obtained in x-space to momentum space.

Expressions for the correlation function at both the hadron and quark-gluon levels have been derived. The subsequent step is to derive the sum rules for the magnetic dipole moment from these expressions. The analytical expressions of the magnetic dipole moments of the $P_{cs}$ state, obtained using three distinct interpolating currents, are presented in the following equations:
\begin{align}
 \mu_{ J^1_{\mu}}& = \frac{e^{\frac{m^2_{P_{cs}}}{\mathrm{M^2}}}}{\lambda^2_{J^1_{\mu}}  }\, \rho_1 (\mathrm{M^2},\mathrm{s_0}),~~~~
\mu_{ J^2_{\mu}}  =\frac{e^{\frac{m^2_{P_{cs}}}{\mathrm{M^2}}}}{\lambda^2_{J^2_{\mu}}}\, \rho_2 (\mathrm{M^2},\mathrm{s_0}),
~~~~
 \mu_{ J^3_{\mu}}  =\frac{e^{\frac{m^2_{P_{cs}}}{\mathrm{M^2}}}}{\lambda^2_{J^3_{\mu}} }\, \rho_3 (\mathrm{M^2},\mathrm{s_0}). 
\end{align}

The analytical expressions of the functions $\rho_i (\mathrm{M^2},\mathrm{s_0})$ are presented in appendix. 

\end{widetext}

\section{Numerical analysis and discussion}\label{numerical}

In this section, a numerical analysis is performed on the QCD light-cone sum rule to predict the magnetic dipole moments of the $P_{cs}$ state.  To perform the numerical analysis of the magnetic dipole moment, it is necessary to determine the numerical values of several parameters. 
The values that have been taken are as follows:  
$m_s(\mu= 2 ~\mbox{GeV}) =93.4^{+8.6}_{-3.4}\,\mbox{MeV}$, $m_c (\mu= m_c) = 1.27 \pm 0.02\,\mbox{GeV}$~\cite{ParticleDataGroup:2022pth},   
$\langle \bar uu\rangle (\mu= 1 ~\mbox{GeV}) = 
\langle \bar dd\rangle (\mu= 1 ~\mbox{GeV})=(-0.24 \pm 0.01)^3\,\mbox{GeV}^3$, $\langle \bar ss\rangle (\mu= 1 ~\mbox{GeV})= (0.8 \pm 0.1)\, \langle \bar uu\rangle$ $\,\mbox{GeV}^3$ \cite{Ioffe:2005ym},
$m_0^{2} (\mu= 1 ~\mbox{GeV}) = 0.8 \pm 0.1 \,\mbox{GeV}^2$ \cite{Ioffe:2005ym},  $\langle g_s^2G^2\rangle (\mu= 1 ~\mbox{GeV}) = 0.48 \pm 0.14~ \mbox{GeV}^4$~\cite{Narison:2018nbv}, $m_{J^1_{\mu}}(\mu= 2.7 ~\mbox{GeV}) = 4.51 \pm 0.12$ GeV,   $m_{J^2_{\mu}}(\mu= 2.7 ~\mbox{GeV}) = 4.51 \pm 0.12$ GeV, $m_{J^3_{\mu}}(\mu= 2.7 ~\mbox{GeV}) = 4.52 \pm 0.12$ GeV,
$\lambda_{J^1_{\mu}} (\mu= 2.7 ~\mbox{GeV}) = (2.75 \pm 0.45) \times 10^{-3} ~\mbox{GeV}^6$, 
$\lambda_{J^2_{\mu}} (\mu= 2.7 ~\mbox{GeV})= (4.64 \pm 0.77) \times 10^{-3} ~\mbox{GeV}^6$, and  $\lambda_{J^3_{\mu}}(\mu= 2.7 ~\mbox{GeV})= (4.64 \pm 0.79) \times 10^{-3} ~\mbox{GeV}^6$~\cite{Wang:2015ixb}. In numerical analysis, we set $m_u$ =$m_d$ = 0, but consider terms proportional to $m_s$.  As is evident, a considerable number of parameters employed in the numerical analysis are dependent on the renormalization scale ($\mu$), with all parameters being calculated on different scales. To ensure a consistent analysis, it is crucial that these parameters be fixed on the same scale. In this analysis, the scale is considered as $\mu=2.7$ GeV, and all scale-dependent parameters used in the analysis have been rescaled to this scale.  To proceed with additional calculations, it is necessary to utilize the photon DAs and their explicit form, along with the requisite numerical parameters, as presented in Ref.~\cite{Ball:2002ps}.

In addition to the aforementioned numerical input parameters, two further parameters are required for our numerical analysis. These are the continuum threshold parameter, denoted by $\mathrm{s_0}$, and the Borel parameter, denoted by $\mathrm{M^2}$.  In an ideal scenario, our numerical analysis would be independent of these parameters. However, this is not feasible in practice. Consequently, we must identify a region of study where the impact of parameter variation on our numerical results is minimal.  
The $\rm{s_0}$ is not a completely random concept; rather, it describes the scale at which the excited states and the continuum begin to become important in the correlation function.  There have been several attempts to formulate methodologies for the estimation of this parameter.
One technique involves varying the parameter within a rational range while observing the emergence of a $\mathrm{M^2}$ window. At this point, the dependence of predictions on the $\mathrm{M^2}$ is known to have ceased~\cite{Ligeti:1993qd}. An alternative technique involves selecting the value of $\rm{s_0}$ as a function of the $\mathrm{M^2}$ and subsequently determining the functional form of the resulting relationship by ensuring that the mass results is unaffected by the $\mathrm{M^2}$~\cite{Lucha:2009uy}. Furthermore, an additional method for optimizing this parameter exists, whereby the continuum threshold parameter can be evaluated independently~\cite{Chen:2015moa, Chen:2016otp}. Each perspective has its own set of advantages and disadvantages with respect to the reliability of the analyses.  Nevertheless, the most common approach to estimate this parameter is to postulate that $\rm{s_0} $ is constrained to the range of $(\rm{m_H}+0.4)\, \rm{GeV }^2 \leq\rm{s_0} \leq (m_H+0.8) \, \rm{GeV }^2$. It is recommended that researchers seeking to determine the $\rm{s_0}$, refer to the experimental data on the mass gaps between the ground states (1S) and the first radial excited states (2S) of pentaquark states. Nonetheless, in light of the absence of experimental data concerning the excited states of the $P_{cs}$ pentaquarks, a similar strategy can be formulated to investigate this phenomenon by examining possible hidden-charm tetraquark candidates. If we prefer the tetraquark state scenarios to other interpretations, we can assign $X(3915)$ and $X(4500)$ as the 1S and 2S hidden-charm tetraquark states, respectively, with $\rm{J^{PC}}=0^{++}$~\cite{Lebed:2016yvr, Wang:2016gxp}, given the possible quantum numbers, decay channels, and mass discrepancies. The $Z_c(3900)$ and $Z_c(4430)$ can be assigned as the 1S and 2S hidden-charm tetraquark states with $\rm{J^{PC}}=1^{+-}$~\cite{Maiani:2014aja, Nielsen:2014mva, Wang:2014vha, Agaev:2017tzv}, respectively. The $Z_c(4020)$ and $Z_c(4600)$ can be assigned as the 1S and 2S hidden-charm tetraquark states with  $\rm{J^{PC}}=1^{+-}$~\cite{Chen:2019osl, Wang:2019hnw}, respectively. Furthermore, the $X(4140)$ and $X(4685)$ can also be assigned as the 1S and 2S hidden-charm tetraquark states with $\rm{J^{PC}}=1^{++}$~\cite{Wang:2021ghk}. As indicated by the aforementioned predictions, it can be seen that the discrepancies in the masses of the 1S and 2S hidden-charm tetraquark states are estimated to lie within the range of approximately $0.5$ to $0.6$ $\,\rm{GeV}$. Therefore, it is reasonable to conclude that a comparable scenario could apply to the excited states of the $P_{cs}$ states. It then turns out that the $\rm{s_0}$ for the $P_{cs}$ pentaquark states can be set to $(\rm{m_H}+0.6)\, \rm{GeV }^2 \leq\rm{s_0} \leq (m_H+0.8) \, \rm{GeV }^2$. 
The working region for the $\mathrm{M^2}$, which is the interval where the variation in our numerical results for this parameter is small, is subject to the constraints imposed by the methodology used. These constraints are referred to as pole contribution (PC) and convergence of OPE (CVG).  The aforementioned constraints are defined and quantified through the use of the corresponding formulae, as follows:
\begin{align}
 \mathrm{PC} &=\frac{\rho (\mathrm{M^2},\mathrm{s_0})}{\rho (\mathrm{M^2},\infty)}, ~~~~
 \mathrm{CVG}=\frac{\rho^{\mathrm{DimN}} (\mathrm{M^2},\mathrm{s_0})}{\rho (\mathrm{M^2},\mathrm{s_0})},
 \end{align}
 where DimN $\geq (8+9+10)$. 
 
\begin{figure}[htp]
\centering
\subfloat[]{\includegraphics[width=0.45\textwidth]{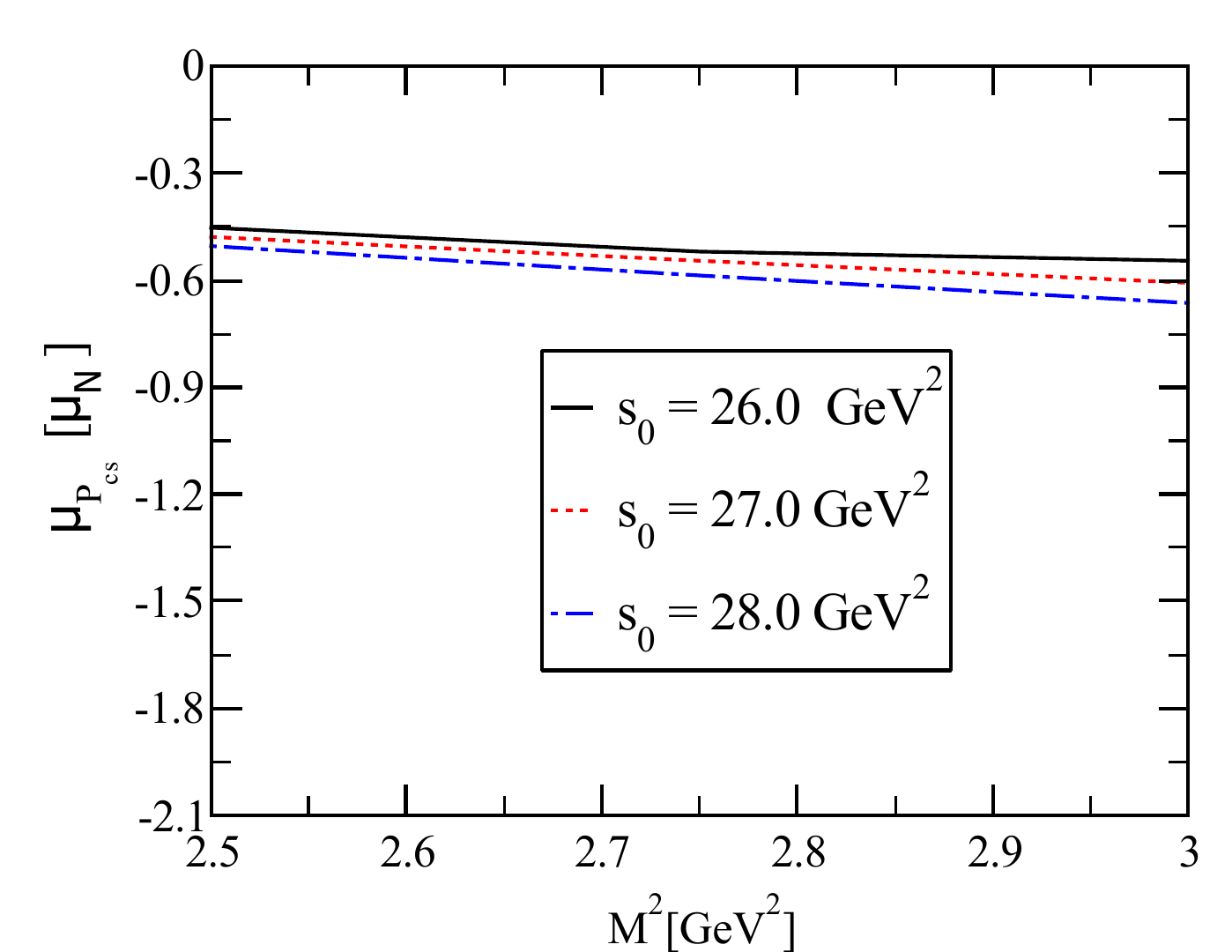}}~~~~~~~~ 
\subfloat[]{\includegraphics[width=0.45\textwidth]{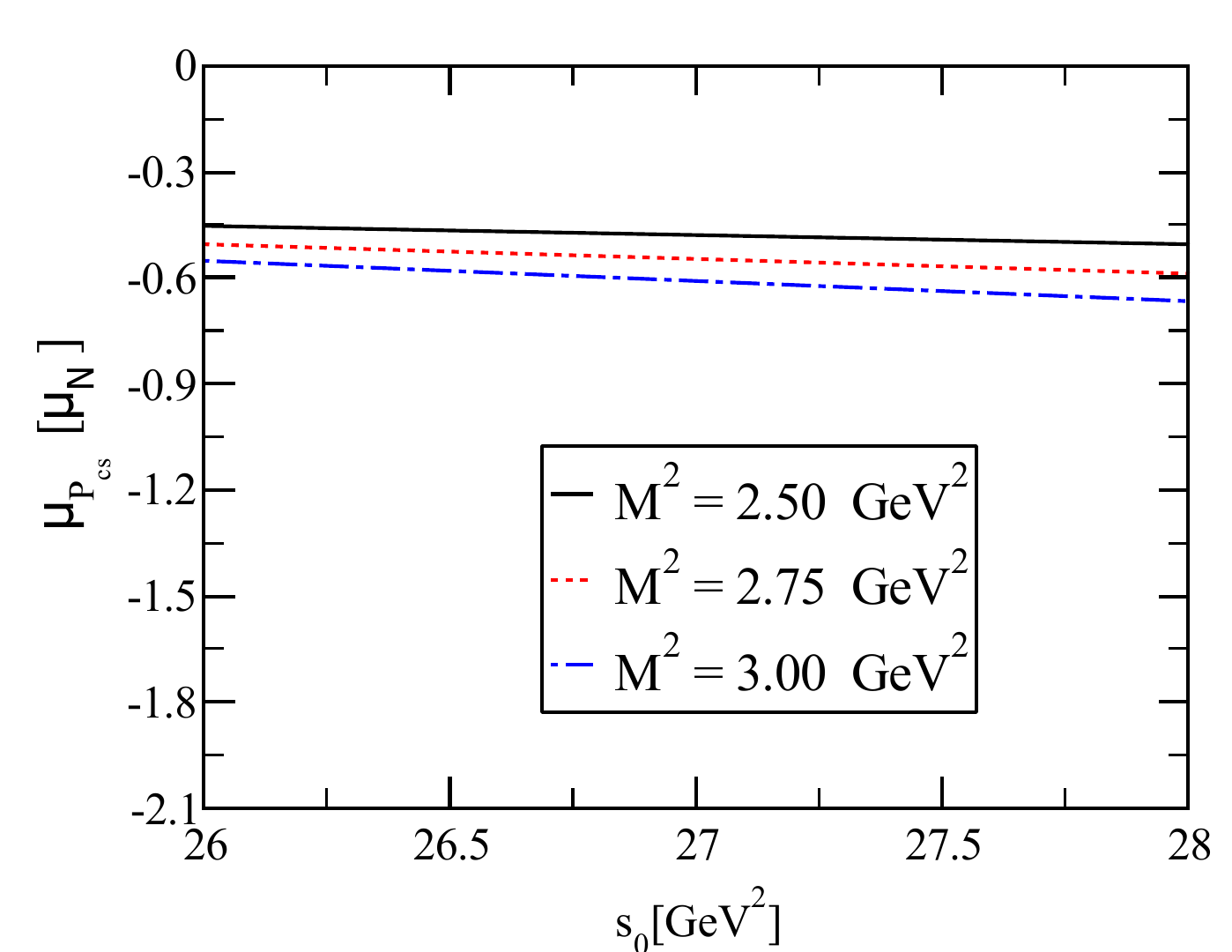}}\\
\subfloat[]{\includegraphics[width=0.45\textwidth]{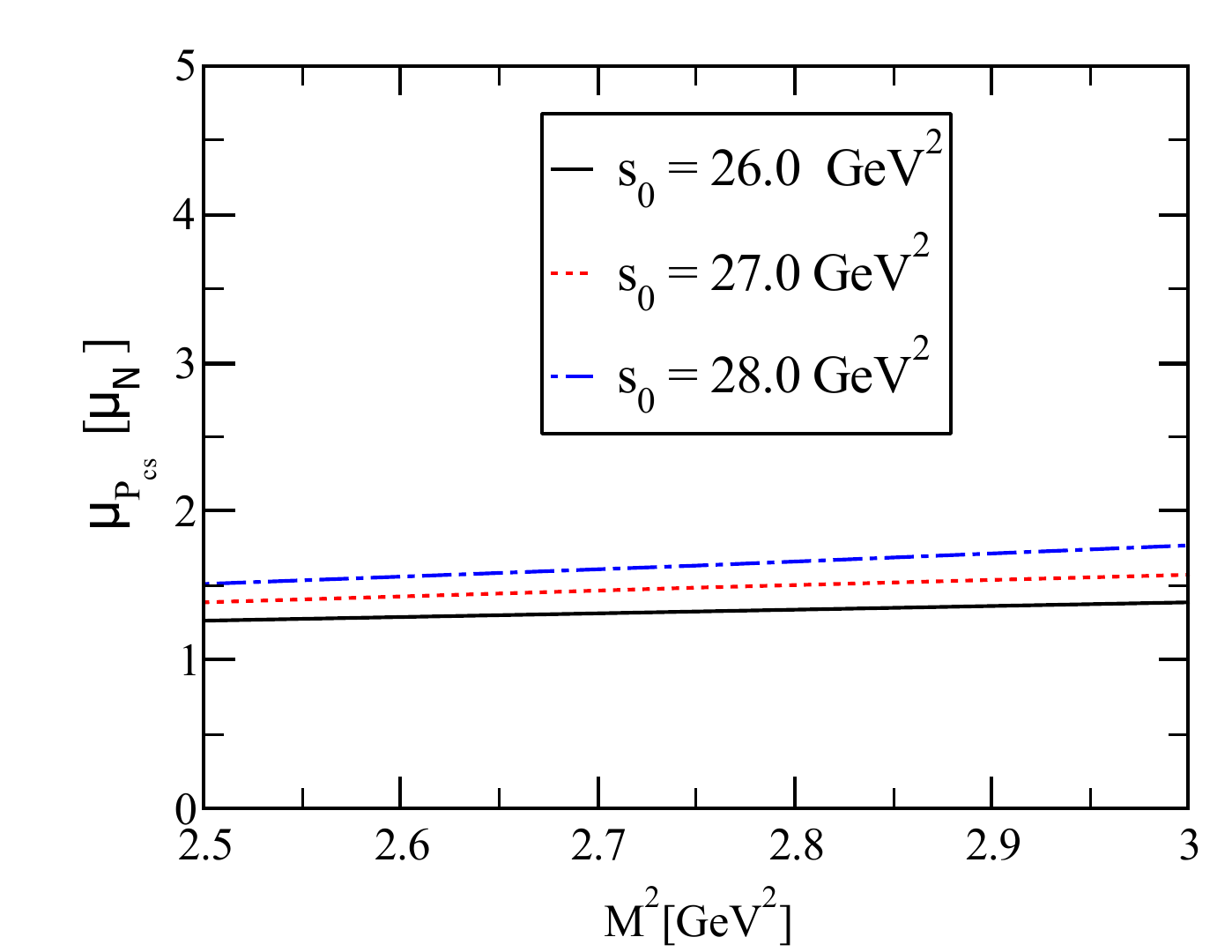}}~~~~~~~~
\subfloat[]{\includegraphics[width=0.45\textwidth]{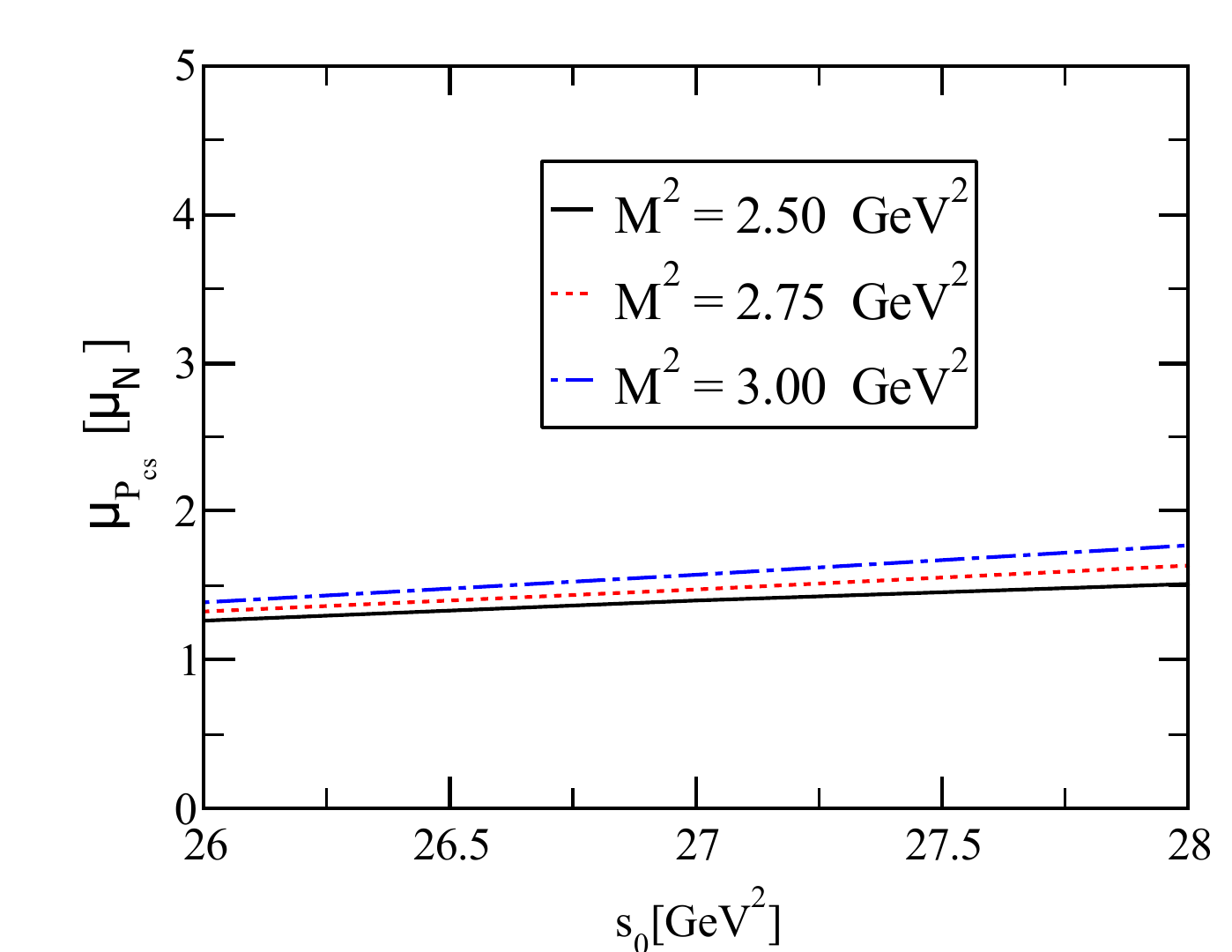}}\\
\subfloat[]{\includegraphics[width=0.45\textwidth]{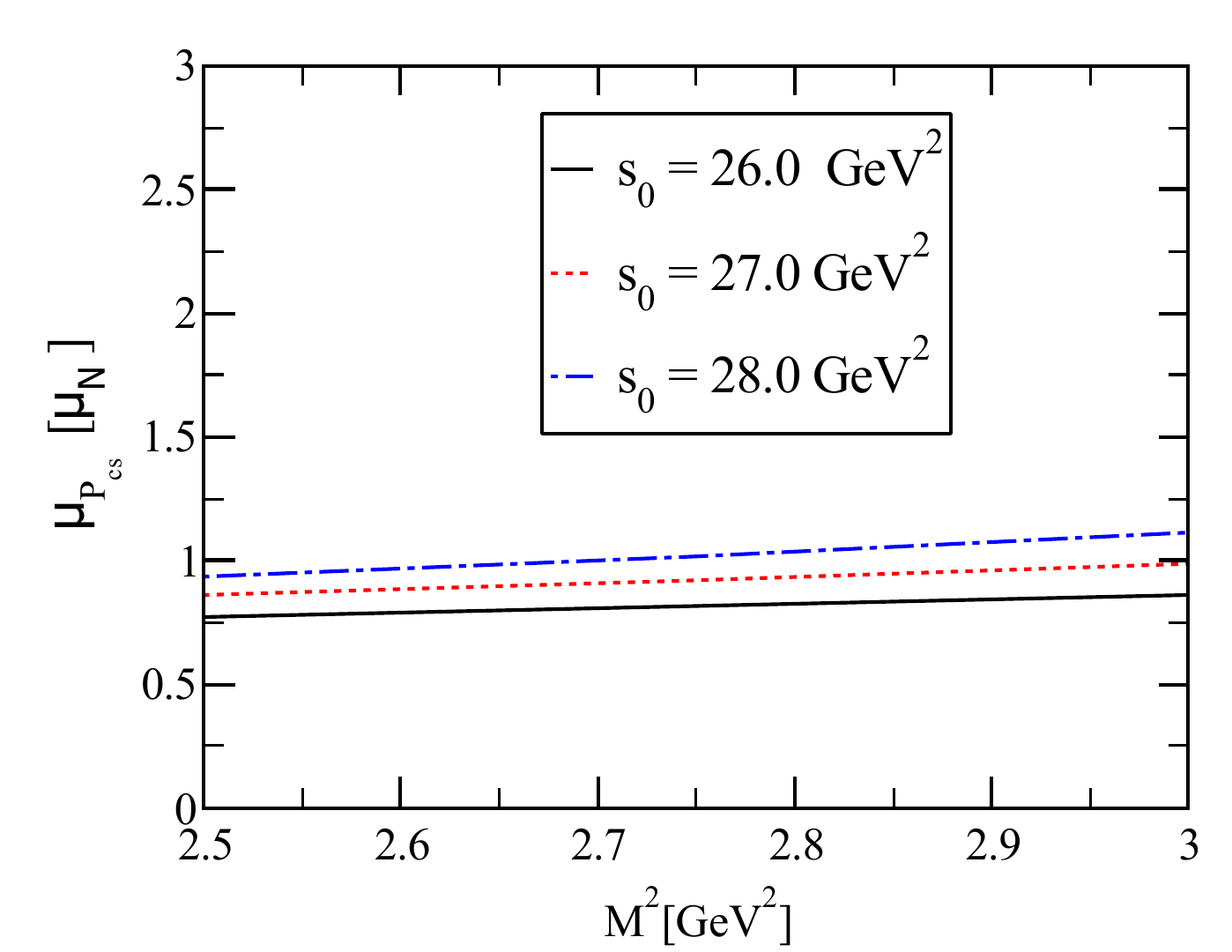}}~~~~~~~~
\subfloat[]{\includegraphics[width=0.45\textwidth]{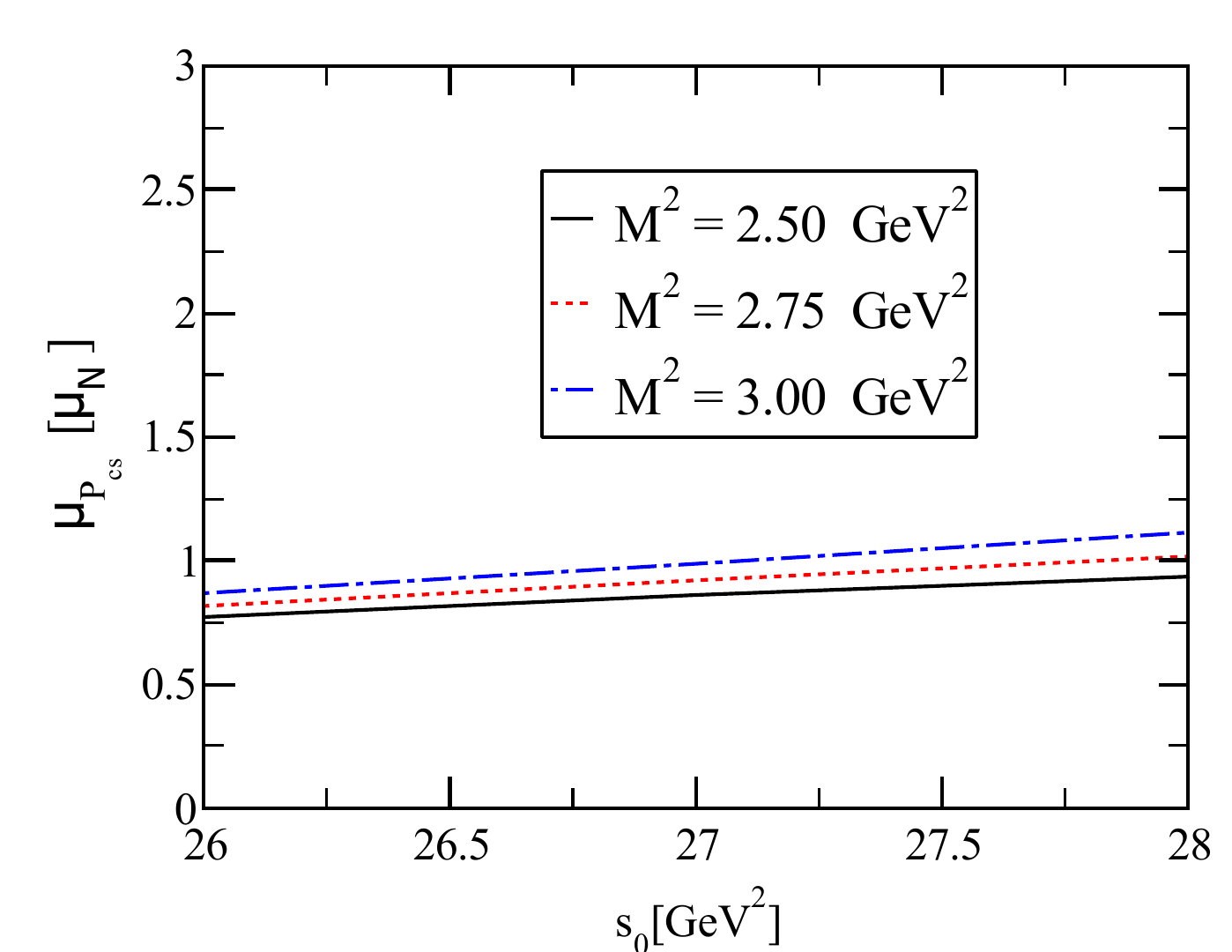}}
 \caption{The magnetic dipole moments of the $P_{cs}$ state versus $\mathrm{M^2}$ (left panel) and $\mathrm{s_0}$ (right panel); (a) and (b)  for the $J^1_\mu$ current; and 
  (c) and (d)  for the $J^2_\mu$ current; and 
 (e)and (f)  for the $J^3_\mu$ current
 .}
 \label{Msqfig}
  \end{figure}

 The PC and CVG values acquired from the computations are presented in Table~\ref{parameter}, along with the working intervals for the state under examination. To ensure the reliability of the obtained working intervals, we have plotted the variation of the magnetic dipole moment results for the aforementioned auxiliary parameters in Fig.~\ref{Msqfig}. The illustration demonstrates a slight variation in the outcomes observed in these intervals, as anticipated.  

\begin{table}[htp]
	\addtolength{\tabcolsep}{10pt}
	\caption{Numerical results of the magnetic dipole moment of the $P_{cs}$ state together with working intervals of $\mathrm{s_0}$ and $\mathrm{M^2}$.}
	\label{parameter}
		\begin{center}
\begin{tabular}{l|ccccc}
	   \hline\hline
	   \\
   State& $\mu~[\mu_N]$& $\mathrm{s_0}$~\mbox{[GeV$^2$]}&$\mathrm{M^2}$~\mbox{[GeV$^2$]}&\mbox{CVG} $[\%]$&\mbox{PC} $[\%]$\\
   \\
\hline\hline
\\
$J_\mu^1$& $ -0.60 \pm 0.15$ &$ [26.0, 28.0]$         &  $ [2.5, 3.0] $& $< 1$ &[60.83, 41.39]\\
\\
$J_\mu^2$ & $ ~~1.60 \pm 0.30$ &$ [26.0, 28.0]$        &  $ [2.5, 3.0]$& $< 1$ &[65.40, 45,31]\\
\\
$J_\mu^3$ & $ ~~0.99 \pm 0.20$ &$ [26.0, 28.0]$        &  $ [2.5, 3.0] $& $< 1$&[66.05, 46.04]\\
\\
	   \hline\hline
\end{tabular}
\end{center}
\end{table}

Now that all the requisite parameters for performing the numerical analysis have been identified, we can proceed to present the resulting numerical values. The complete set of numerical results, including all estimated uncertainty due to the inherent variability of the input parameters, is presented in Table \ref{parameter}. 
It should be emphasized that adopting a renormalization scale of $\mu = 1~\text{GeV}$ leads to a variation of approximately $(18\text{--}22)\%$ in the electromagnetic multipole moment results, which corresponds to an increase in magnitude when evaluated in absolute terms. The obtained deviations are within the uncertainty associated with the QCD light-cone sum rules method.  We would like to point out that roughly 13$\%$ of the errors in the numerical results are due to the mass of pentaquarks, 30$\%$ belongs to the residue of pentaquarks, 12$\%$ belongs to the mass of the quarks, 24$\%$ belongs to $\rm{s_0}$, 8$\%$ belongs to $\rm{M^2}$, 5$\%$ belongs to photon DAs,
3.4$\%$ belongs to $\langle \bar qq\rangle$, 1.7$\%$ belongs to $\langle \bar ss\rangle$, 1.4$\%$ belongs to $\langle g_s^2G^2\rangle$, and the remaining 1.5$\%$ corresponds to other higher-dimension condensates (such as $\langle \bar qq\rangle \langle \bar ss\rangle \langle g_s^2G^2\rangle$, $\langle \bar qq\rangle^2  \langle g_s^2G^2\rangle$, $\langle \bar qq\rangle  \langle g_s^2G^2\rangle$,
$\langle \bar ss \rangle   \langle g_s^2G^2\rangle$, and $\langle \bar qq\rangle \langle \bar ss\rangle $).

 As illustrated in Table~\ref{parameter}, the distinct interpolating currents employed to investigate the magnetic dipole moments of the $P_{cs}$ pentaquark, which is composed of the same quarks, yield significant discrepancies in the derived values.  It is feasible that the aforementioned phenomenon could be interpreted as a manifestation of more than one $P_{cs}$ pentaquark state, exhibiting identical quantum numbers and a similar composition of quarks, yet differing in their magnetic dipole moments. As previously highlighted, the interpolating currents in question possess identical quantum numbers, thereby resulting in nearly degenerate masses for the aforementioned $P_{cs}$ pentaquark~\cite{Wang:2015ixb}. However, it is evident that the outcomes yielded for magnetic dipole moments exhibit considerable sensitivity to the diquark-diquark-antiquark configuration and the intrinsic characteristics of the state under examination.  It is generally assumed that a change in the basis of hadrons would have no impact on the resulting data. Nevertheless, it is feasible that this assumption may not be applicable in the context of magnetic dipole moments. This is because the magnetic dipole moments of the states under consideration are directly correlated with their internal structural organization. In the language of magnetic dipole moments, a change in the basis of the related state also leads to a change in the internal structure of the hadron, which in turn may result in a significant change in the calculated results. In Refs.~\cite{Ozdem:2024dbq, Azizi:2023gzv, Ozdem:2024rqx, Ozdem:2022iqk}, a variety of interpolating currents have been employed to obtain the magnetic dipole moment of tetra- and pentaquarks. The findings of these studies revealed notable discrepancies in the magnetic dipole moments observed when employing diquark-antidiquark or diquark-diquark-antiquark structures.  Therefore, the selection of disparate interpolating currents that may couple to the same states, or the modification of the isospin and charge basis of the states under examination, may result in disparate magnetic dipole moments. 

To provide a more comprehensive examination, we have also analyzed the individual quark contributions, the results of which are presented in Table \ref{parameter2}. It is important to note that the results presented here are based on the central value of all input parameters. As is seen, the magnetic dipole moments of light-quarks cancel each other out and the magnetic dipole moment results consist only of the charm-quark contribution. Moreover, the table demonstrates the influence of disparate diquark configurations on the magnetic dipole moment results.
\begin{table}[htp]
	\addtolength{\tabcolsep}{10pt}
	\caption{Individual quark contributions to the magnetic dipole moments.}
	\label{parameter2}
		\begin{center}
\begin{tabular}{l|ccccc}
	   \hline\hline
	   \\
   State& $\mu_u~[\mu_N]$& $\mu_d~[\mu_N]$&$\mu_s~[\mu_N]$&$\mu_c~[\mu_N]$&$\mu_{total}~[\mu_N]$\\
   \\
\hline\hline
\\
$J_\mu^1$& $~~0.004$ &$ -0.002$ &  $ -0.002$ & $-0.60$ &$-0.60$\\
\\
$J_\mu^2$ & $ -0.030$ &$ ~~0.015$  &  $~~0.015$& $~~1.60$ &$~~1.60$\\
\\
$J_\mu^3$ & $ ~~0.220$ &$ -0.110$  &  $-0.110$& $~~0.99$&$~~0.99$\\
\\
	   \hline\hline
\end{tabular}
\end{center}
\end{table}

To obtain further insight, it would be useful to conduct a comparative analysis between the numerical values obtained and the existing literature on the subject.  In Refs.~\cite{Li:2021ryu, Gao:2021hmv}, the magnetic dipole moments of the $P_{cs}$ state have been investigated employing the quark model in the molecular  configuration with the quantum numbers $J^P = \frac{3}{2}^-$. The resulting values are $\mu_{P_{cs}} = 0.465~\mu_N$ and $\mu_{P_{cs}} = - 0.231~\mu_N$, respectively.   In Ref.~\cite{Ozdem:2022kei}, the magnetic dipole moment of the pentaquark $P_{cs}$ was determined in the context of the QCD light-cone sum rules method. This was achieved by considering within a molecular framework, wherein it was assumed that the particle in question exhibited quantum numbers $J^P = \frac{3}{2}^-$. The resulting value was subsequently obtained, and it was found to be $\mu_{P_{cs}} = - 1.67 \pm 0.58 ~\mu_N$. In Refs. \cite{Ozdem:2021ugy,Gao:2021hmv}, the magnetic dipole moment of the $P_{cs}$ state is also calculated in the framework of the QCD light-cone sum rules methods and quark model  within the diquark-diquark-antiquark framework. The calculation is performed under the assumption that this state has $J^P = \frac{1}{2}^-$ quantum numbers. The results of this calculation are $\mu_{P_{cs}} =0.233~\mu_N$ and $\mu_{P_{cs}} = 0.34^{+0.13}_{-0.11}~\mu_N$. In Refs.~\cite{Ozdem:2021ugy,Li:2021ryu, Gao:2021hmv}, the magnetic dipole moments of the $P_{cs}$ state have also been calculated using the QCD light-cone sum rules methods and quark model in the molecular  configuration with the quantum numbers $J^P = \frac{1}{2}^-$. The resulting values are $\mu_{P_{cs}} = 1.75^{+0.64}_{-0.58}~\mu_N$, $\mu_{P_{cs}} = -0.062~\mu_N$ and $\mu_{P_{cs}} = - 0.531~\mu_N$, respectively. The results exhibit considerable discrepancies, not only in magnitude but also in sign. To facilitate comparison of the results obtained using different models and approaches, and to enhance their intelligibility, they are presented in Table \ref{comp1}. 
The disparity in the results may be related to the use of various models and interactions, as well as to the choice of different parameter sets.
The discrepancy in the findings derived from quark models can be more straightforwardly interpreted by considering the different quark masses and wave functions employed. However, a discrepancy in the findings derived from the molecular and compact diquark frameworks, employing QCD light-cone sum rules, demands further investigation.  In order to accomplish this task, it is recommended to examine the contributions of individual quarks. A thorough analysis of these contributions reveals that light quarks significantly impact the molecular frame's results, leading to dominant outcomes. Conversely, within the compact diquark structure, as illustrated in Table \ref{parameter2}, the contributions of light quarks are neutralized, resulting in a c-quark-dominated magnetic dipole moment. The findings reveal that the diquark structures selected during the hadron construction process exert a direct influence on the results obtained.
From both the results in the literature and the results obtained in this study, it can be concluded that the magnetic dipole moments of the $P_{cs}$ state are capable of projecting its inner structure, which can be used to determine its quantum numbers and quark-gluon organization.  
Nevertheless, it is recommended that further studies be conducted using alternative non-perturbative techniques to gain a more comprehensive understanding of the observed results.

\begin{table}[htp]
	\addtolength{\tabcolsep}{10pt}
	\caption{Our results and other theoretical results for the magnetic dipole moments of $P_{cs}$ state (in units of  $\mu_N$).}
	\label{comp1}
		\begin{center}
\begin{tabular}{l|ccccc}
	   \hline\hline
	   \\
   Approaches & $\mathrm{J^P}$& Compact Diquark&Molecule\\
   \\
\hline\hline
\\
\cite{Ozdem:2021ugy}&  &$0.34^{+0.13}_{-0.11}$ &  $1.75^{+0.64}_{-0.58}$ \\
\\
\cite{Li:2021ryu}& $\rm{\frac{1}{2}^-}$ &$-$ &  $ -0.062$ \\
\\
\cite{Gao:2021hmv}&  &$ 0.223$ &  $ -0.531$ \\
\\
\hline\hline
\\
\cite{Li:2021ryu}&  &$ -$ &  $ 0.465$ \\
\\
\cite{Ozdem:2022kei}&  &$-$ &  $-1.67 \pm 0.58$ \\
\\
\cite{Gao:2021hmv}& $\rm{\frac{3}{2}^-}$ &$ -0.231$ &  $ -0.231$ \\
\\
This Work  [$J_\mu^1$]&  &$-0.60 \pm 0.15$ &  $-$ \\
\\
This Work  [$J_\mu^2$]&  &$~~1.60 \pm 0.30$ &  $-$ \\
\\
This Work  [$J_\mu^3$]&  &$~~0.99 \pm 0.20$ &  $-$ \\
\\
	   \hline\hline
\end{tabular}
\end{center}
\end{table}

 To ensure a comprehensive analysis, the higher multipole moments, i.e. the electric quadrupole ($\mathcal Q$) and magnetic octupole ($\mathcal O$) moments, of this pentaquark, are also determined, and the results are presented as follows:
%
 \begin{align}
  \mathcal Q_{{J_\mu^1}}&= (-0.20 \pm 0.02 ) \times 10^{-2}~\mbox{fm}^2 ,~~~~~~
  \mathcal O_{{J_\mu^1}}= (-0.032 \pm 0.003) \times 10^{-3}~\mbox{fm}^3,
  \end{align}
  \begin{align}
  \mathcal Q_{{J_\mu^2}}&=(0.74 \pm 0.07) \times 10^{-2}~\mbox{fm}^2 ,~~~~~~~
  \mathcal O_{{J_\mu^2}}=(-0.35 \pm 0.03) \times 10^{-3}~\mbox{fm}^3, 
  \end{align}
  \begin{align}
  \mathcal Q_{{J_\mu^3}}&= (0.48 \pm 0.06) \times 10^{-2}~\mbox{fm}^2 ,~~~~~~~ 
  \mathcal O_{{J_\mu^3}}= (-0.22 \pm 0.03) \times 10^{-3}~\mbox{fm}^3 .
 \end{align}
 
As can be observed from the presented results, the higher multipole moments, like the magnetic dipole moment results, demonstrate considerable variability about the diquark-diquark-antiquark structure chosen for the hadron in question.  It can be observed that the magnitudes of the electric quadrupole and magnetic octupole moments are considerably less than that of the magnetic dipole moment. The outcomes for the higher multipole moments yielded values that are inconsistent with zero, indicating that the charge distribution is not spherical.  It is well established that the magnitude of the higher multipole moments provides information about the deformation of the associated hadron and its direction. In the case of the $J_\mu^1$ current, it is found that both the electric quadrupole and magnetic octupole moments have non-zero values and negative signs, suggesting that the quadrupole and octupole moment distributions of this state are oblate and have the same geometric shape as the charge distribution. In the case of the $J^2_\mu$ and $J^3_\mu$ currents, the higher multipole moments are found to have non-zero values and positive signs for the quadrupole moment and negative signs for the octupole moment. This indicates that the quadrupole moment distributions are prolate and that the charge distribution and geometrical shape are opposite.

 \section{Summary}\label{summary}

In this study, a comprehensive calculation of the magnetic dipole moment of the $P_{cs}(4459)$ pentaquark has been conducted within the domain of the QCD light-cone sum rules method. In the conducted analysis, given that this pentaquark has $J^P =\frac{3}{2}^-$ quantum numbers and a compact pentaquark structure, three possible interpolating currents that are likely to couple this state have been considered. As can be seen from the results in Table \ref{parameter}, the different interpolating currents used to probe the magnetic dipole moments of $P_{cs}(4459)$ pentaquarks consisting of the same quarks lead to significant differences in the obtained values.  It is possible to interpret the above phenomenon as a consequence of the existence of more than one $P_{cs}(4459)$ pentaquark, which shows the same quantum numbers and a similar quark combination but has different magnetic dipole moments. The numerical results obtained have led to the conclusion that the magnetic dipole moments of the $P_{cs}(4459)$ state are capable of projecting its inner structure, which can be used to determine its quantum numbers and quark-gluon organization. Our findings for the magnetic dipole moment of the $P_{cs}(4459)$ state are also compared with the results of the other theoretical models existing in the literature. The results exhibit considerable discrepancies, not only in magnitude but also in sign. The discrepancy in the results may be related to the use of various models and interactions, as well as to the choice of different parameter sets. To get a more conclusive picture of these results, further studies are encouraged. Additionally, the higher multipole moments, the electric quadrupole, and magnetic octupole moments are calculated for this state. It is observed that the results for the higher multipole moments are affected by the choice of interpolating currents, like that observed for the magnetic dipole moment results.

An understanding of the structural characteristics and the ability to identify them in a photo-production process would be facilitated by the study of the magnetic dipole moments of the hidden-charm pentaquarks with and without strangeness. This approach could provide an independent probe of pentaquarks. We hope that our determinations of the magnetic dipole moment of the $P_{cs}(4459)$ pentaquark when considered alongside the findings of other theoretical investigations into the spectroscopic parameters and decay widths of this pentaquark, will prove useful in future experiments designed to investigate these subjects and to elucidate the internal structure of the $P_{cs}(4459)$ pentaquark.

 \newpage
 
\newpage 
 \section*{Appendix: Sum rules for the $\rho_i (\mathrm{M^2},\mathrm{s_0})$ functions} \label{appb}
  
  This appendix presents the analytical results of the magnetic dipole moment analysis of the $P_{cs}$ state for the interpolating currents $J_{\mu}^1$, $J_{\mu}^2$ and $J_{\mu}^3$.
 
 \subsection*{1-The obtained sum rule for the magnetic dipole moment
of the $J_{\mu}^1$ interpolating current}
  \begin{align}
  \rho_1 (\mathrm{M^2},\mathrm{s_0}) &=   F_1^{J^1_{\mu}}(\mathrm{M^2},\mathrm{s_0}) - \frac{1}{m_{J_\mu^1}} F_2^{J^1_{\mu}}(\mathrm{M^2},\mathrm{s_0}),
 \end{align}
 where, 
\begin{align}
 F_1^{J^1_{\mu}}(\mathrm{M^2},\mathrm{s_0})&=  - \frac{e_c}{2^{26}\times 3^3 \times 5^3 \times 7 ^3 \pi^7} \Big[1876 m_c m_s I[0, 6] - 855 I[0, 7]\Big]
 \nonumber\\
 &+ \frac{C_1 C_2 C_3}{2^{19}\times 3^6 \times 5  \pi^3} \Big[ 90 e_c m_c m_s I[0, 1] - 11 (2 e_c + e_d + e_u) I[0, 2] \Big]
 \nonumber\\
 &- \frac{11 C_1 C_2^2}{2^{19}\times 3^6 \times 5  \pi^3}(e_c + e_s) I[0, 2]
 \nonumber\\
 &+ \frac{11 C_1 C_2 m_0^2}{2^{23}\times 3^7 \times 5  \pi^5}  \Big[44 e_c m_c - 112 m_c (e_d + 2 e_s + e_u)  I[0, 2] + 
   33  m_s (2 e_c + e_d + e_u)  I[0, 2]\Big]
   \nonumber\\
 &+ \frac{11 C_1 C_3 m_0^2}{2^{23}\times 3^6 \times 5  \pi^5}\Big[
 \Big(-7 (e_d + e_u) (48 m_c + 11 m_s) + 22 e_c (3 m_c + 16 m_s)\Big) I[0, 2]\Big]
  \nonumber\\
 &+ \frac{ C_1 C_2}{2^{22}\times 3^7 \times 5  \pi^5}\Big[
 \Big(75 m_c (e_d + 2 e_s + e_u) + 202 m_s (e_d + e_u)  - 
   e_c (443 m_c + 476 m_s)\Big) I[0, 3] \Big]
   \Big]
  \nonumber\\
 &+ \frac{ C_1 C_3}{2^{24}\times 3^7 \times 5  \pi^5}\Big[
 \Big(6 (e_d + e_u) (50 m_c + 7 m_s) + e_c (-886 m_c + 129 m_s)\Big) I[0, 3] \Big]
 \nonumber\\
 &+ \frac{e_c C_2 C_3}{2^{15}\times 3^4 \times 5  \pi^3}\Big[2 m_c m_s I[0, 3] + 3 I[0, 4] \Big]
 \nonumber\\
 &+ \frac{ C_1}{2^{28}\times 3^7 \times 5^2  \pi^7}\Big[ 
 30 m_c m_s (273 e_c - 10 (e_d + e_u))  I[0, 4] - 
  (363 e_c + 1148 (e_d + es + e_u)) I[0, 5]\Big]
  \nonumber\\
 &+ \frac{7e_c  C_2}{2^{20}\times 3^5 \times 5^2  \pi^5}\Big[ 
(8 m_c - 27 m_s) I[0, 5]\Big]
\nonumber\\
 &+ \frac{e_c  C_3}{2^{20}\times 3^5 \times 5^2  \pi^5}\Big[ 
(28 m_c + 81 m_s) I[0, 5]\Big],\\
\nonumber\\
%
 F_2^{J^1_{\mu}}(\mathrm{M^2},\mathrm{s_0})&= \frac{e_c \, m_c}{2^{26}\times 3^5 \times 5^3 \times 7^2 \pi^7} \Big [41440 m_c m_s I[0, 6] - 18063 I[0, 7]\Big]\nonumber\\
 &+\frac{C_1 C_2 C_3 \, m_c}{2^{18}\times 3^6  \pi^3} \Big[-2(e_d+e_u)\big(3 m_c m_s I[0, 1] - 2 I[0, 2] + 7 I[1, 1]\big) + 
 e_c \big(36 m_c m_s I[0, 1] - 11 I[0, 2]\nonumber\\
 &+ 10 I[1, 1]\big)\Big]\nonumber\\
 &-\frac{ C_1 C_2^2 \, m_c}{2^{19}\times 3^6  \pi^3}\Big[ (11 e_c - 8 e_s) I[0, 2] + 2 (-5 e_c + 14 e_s) I[1, 1]\Big]\nonumber\\
 &+ \frac{C_1 C_2 m_0^2\,m_c}{2^{23}\times 3^3 \times 5  \pi^5}
 \Big[\Big (11  m_c (e_d + 2 e_s + e_u) + 8 m_s (e_d + e_u) - 
    2 e_c (52 m_c + 11 m_s)\Big) I[0, 2] + 
 2 \Big (72 e_c m_c \nonumber\\
 &- 19 m_c (e_d + 2 e_s + e_u) + 10 e_c m_s - 
    14 m_s (e_d + e_u) \big) I[1, 1]\Big]
   \nonumber\\
    &+ \frac{C_1 C_3 m_0^2\,m_c}{2^{23}\times 3^6   \pi^5}
 \Big[2 e_c (-78 m_c + 37 m_s) I[0, 2] + 
 4 e_c (54 m_c - 5 m_s) I[1, 
   1] + (e_d + e_u) \Big((33 m_c - 28 m_s) I[0, 2] \nonumber\\
   &+ 
    2 (-57 m_c + 14 m_s) I[1, 1]\Big) \Big]
 \nonumber
 \end{align}
  \begin{align}
  &+ \frac{C_1 C_2\,m_c}{2^{21}\times 3^7\pi^5}
 \Big[ \Big (20 m_c (e_d + 2 e_s + e_u) + e_c (103 m_c - 112 m_s) + 
    26 m_s (e_d + e_u) \Big) I[0, 3] + 
 3 \Big (29 m_c (e_d + 2 e_s \nonumber\\
 &+ e_u) + 28 m_s (e_d + e_u) - 
    10 e_c (9 m_c + 2 m_s)\Big) I[1, 2] \Big]
 \nonumber\\
 &+ \frac{C_1 C_3 \,m_c}{2^{23}\times 3^7 \times 5   \pi^5}
 \Big[e_c (1030 m_c + 71 m_s) I[0, 3] + 300 e_c (-9 m_c + 2 m_s) I[1, 2] + 
 2 (e_d + e_u) \Big (8 (25 m_c \nonumber\\
 &- 7 m_s) I[0, 3] + 
    3 (290 m_c - 77 m_s) I[1, 2]\Big) \Big]
 \nonumber\\
 & -\frac{C_2 C_3 \,e_c \,m_c}{2^{14}\times 3^5 \times 5 \pi^5} \Big[ 10 m_c m_s I[0, 3] + 9 I[0, 4]\Big]
  \nonumber\\
 & +\frac{C_2^2 \, e_c\, m_c}{2^{15}\times 3^3 \times 5 \pi^5} \Big[ 20 m_c m_s I[0, 3] - I[0, 4] \Big]
 \nonumber\\
 & +\frac{C_1\, m_c}{2^{30}\times 3^6 \times 5^2 \pi^7} \Big[ 40 \big (-320  m_c m_s (e_d + e_u) I[0, 4] - 
    7 (e_d + e_s + e_u) I[0, 5] - 
    784  m_c m_s (e_d + e_u) I[1, 3]\big) \nonumber\\
    &+ 
 e_c \big (13143 I[0, 5] + 
    1120 m_c m_s (-11 I[0, 4] + 36 I[1, 3])\big) + 
 900  (e_d + e_s + e_u) I[1, 4]\Big]\nonumber\\
 & -\frac{C_2\,e_c\, m_c}{2^{19}\times 3^4 \times 5^2 \times  \pi^5} \Big[23 m_c - 72 m_s \Big] I[0, 5] \nonumber\\
 &-
 \frac{C_3\,e_c\, m_c}{2^{22}\times 3^4 \times 5^2 \times  \pi^5} \Big[92 m_c + 243 m_s\Big] I[0, 5],
\end{align}

\subsection*{2-The obtained sum rule for the magnetic dipole moment
of the $J_{\mu}^2$ interpolating current}
  \begin{align}
   \rho_2 (\mathrm{M^2},\mathrm{s_0}) &=   F_1^{J^2_{\mu}}(\mathrm{M^2},\mathrm{s_0}) - \frac{1}{m_{J_\mu^2}} F_2^{J^2_{\mu}}(\mathrm{M^2},\mathrm{s_0}),
 \end{align}
 where,
\begin{align}
  F_1^{J^2_{\mu}}(\mathrm{M^2},\mathrm{s_0})&=  \frac{1}{2^{25}\times 3^3 \times 5^3 \times 7 ^3 \pi^7} \Big[
  6559 e_c m_c m_s I[0, 6] - 45 (19 e_c + 26 e_d + 26e_s + 26 e_u) I[0, 7]\Big]
 \nonumber\\
 &- \frac{C_1 C_2 C_3}{2^{19}\times 3^5 \times 5  \pi^3} \Big[  (4 e_c + 3 e_d + 3e_u) (30 m_c m_s I[0, 1] - I [0, 2])  \Big]
 \nonumber\\
 &+ \frac{11 C_1 C_2 m_0^2}{2^{23}\times 3^6 \times 5  \pi^5}  \Big[ \Big(5 m_c (239 e_d + 3358 e_s + 239 e_u)  + 594 m_s(e_d + e_u) + 
   8 e_c (-628 m_c + 99 m_s)\Big) I[0, 2] \Big]
   \nonumber\\
 &- \frac{11 C_1 C_3 m_0^2}{2^{23}\times 3^7 \times 5  \pi^5}\Big[  \Big((e_d + e_u) (219 m_c + 352 m_s) + 3 e_c (896 m_c + 429 m_s)\Big) [0, 2] \Big]
  \nonumber\\
 &+ \frac{ C_1 C_2}{2^{21}\times 3^7 \times 5  \pi^5}\Big[
 \Big(1855 e_c m_c - 12 m_c (20 e_d + 463 e_s + 20 e_u)  + 603 e_c m_s + 
   136m_s (e_d + e_u) \Big) I[0, 3]
   \Big]
  \nonumber\\
 &- \frac{ C_1 C_3}{2^{23}\times 3^7 \times 5  \pi^5}\Big[ \Big(15 (e_d + e_u) (76 m_c + 5 m_s) - e_c (4304 m_c + 338 m_s)\Big) I[0, 3]\Big]
 \nonumber\\
 &+ \frac{e_c C_2 C_3}{2^{14}\times 3^4 \times 5  \pi^3}\Big[ 2 m_c m_s I[0, 3] + 3 I[0, 4]\Big]
 \nonumber\\
  &+\frac{e_c \,C_2^2}{2^{15}\times 3^4 \times 5  \pi^3} \Big[ 116 m_c m_s I[0, 3] - 3 I[0, 4] \Big]\nonumber\\
 &+ \frac{ C_1}{2^{29}\times 3^7 \times 5^2  \pi^7}\Big[   105 m_c m_s  (-8812 e_c + 4205 (e_d + e_u))  I[0, 4] - 
 32  (1953 e_c + 1891 e_d - 9278 e_s \nonumber\\
 &+ 1891 e_u) I[0, 5]\Big]
  \nonumber\\
 &+ \frac{  C_2}{2^{21}\times 3^5 \times 5^2  \pi^5}\Big[  \Big(-5312 e_c m_c + 5292 e_c m_s + 4131 (e_d + e_u) m_s \Big) I[0, 5]\Big]
\nonumber\\
 &- \frac{e_c  C_3}{2^{22}\times 3^4 \times 5^2  \pi^5}\Big[ 
 \Big(664 e_c m_c + 81 m_s (7 e_c + e_d + e_u) \Big) I[0, 5]\Big],\\
 \nonumber
\end{align}

\begin{align}
  F_2^{J^2_{\mu}}(\mathrm{M^2},\mathrm{s_0})&= \frac{ m_c}{2^{25}\times 3^4 \times 5^3 \times 7^2 \pi^7} \Big [   4327 e_c I[0, 7] + 28742 e_c I[1, 6] - 
 18 (e_d + e_s + e_u) (88 I[0, 7] + 427 I[1, 6]) \Big]\nonumber\\
 &+\frac{C_1 C_2 C_3 \, m_c}{2^{18}\times 3^6  \pi^3} \Big[
 e_c \big(-208 m_c m_s I[0, 1] + 137 I[0, 2] - 220 I[1, 1]\big) + 
 3 (e_d + e_u) (17 I[0, 2] - 16 I[1, 1])\Big]\nonumber\\
 &+\frac{ C_1 C_2^2 \, m_c}{2^{18}\times 3^6  \pi^3}\Big[
 (37 e_c - 246 e_s) I[0, 2] + 2 (-55 e_c + 138 e_s) I[1, 1]\Big]\nonumber\\
&+ \frac{C_1 C_2 m_0^2\,m_c}{2^{23}\times 3^5  \pi^5}
 \Big[ \Big(m_c(250 e_c - 19 e_d - 1910 e_s - 19 e_u) + 274 e_c m_s + 
    102 m_s(e_d + e_u)  \Big) I[0, 2] - 
 4 \Big(m_c(180 e_c \nonumber\\
 &+ 2 e_d - 905 e_s + 2 e_u) + 110 e_c m_s + 
    24 m_s (e_d + e_u) \Big) I[1, 1]\Big]
   \nonumber\\
    &+ \frac{C_1 C_3 m_0^2\,m_c}{2^{22}\times 3^6   \pi^5}
 \Big[  6 m_c \Big(e_c (56 m_c - 33 m_s) + 2 (e_d + e_u) (19 m_c - 4 m_s)\Big) I[0, 2] +
  m_c \Big(-432 e_c m_c  \nonumber\\
  &- 201 m_c(e_d + e_u)  + 188 e_c m_s + 
    48 m_s (e_d + e_u) \Big) I[1, 1]\Big]
 \nonumber\\
 &+ \frac{C_1 C_2\,m_c}{2^{21}\times 3^7\pi^5}
 \Big[ \Big (m_c (737 e_c + 50 e_d - 746 e_s + 50 e_u) + 124 e_c m_s - 
    60  m_s (e_d + e_u) \Big) I[0, 3] + 
 6 \Big (m_c (225 e_c \nonumber\\
 &+ 4 e_d - 1162 e_s + 4 e_u) + 220 e_c m_s + 
    48 m_s (e_d + e_u)  \Big) I[1, 2] \Big]
 \nonumber\\
 &+ \frac{C_1 C_3 \,m_c}{2^{23}\times 3^7 \times 5   \pi^5}
 \Big[ 2 \Big (-(e_d + e_u) (1030 m_c + 587 m_s) + 
    e_c (2290 m_c + 876 m_s)\Big) I[0, 3] + 
 3 \Big (8 e_c (450 m_c \nonumber\\
 &- 517 m_s) + (e_d + 
       e_u) (2680 m_c - 133 m_s)\Big) I[1, 2] \Big]
 \nonumber\\
 & +\frac{C_2 C_3 \,m_c}{2^{15}\times 3^5 \times 5 \pi^5} \Big[  9 (e_d + e_u) (I[0, 4] - 4 I[1, 3]) + 
 e_c (40 m_c m_s I[0, 3] - 30 I[0, 4] + 352 I[1, 3]) \Big]
  \nonumber\\
 & +\frac{C_2^2 \, m_c}{2^{15}\times 3^5 \times 5 \pi^5} \Big[ -15 e_c (24 m_c m_s I[0, 3] + I[0, 4]) + 
 9 e_s (I[0, 4] - 4 I[1, 3]) + 176 e_c I[1, 3]\Big]
 \nonumber\\
 & +\frac{C_1\, m_c}{2^{30}\times 3^6 \times 5^2 \pi^7} \Big[ -9920  m_c m_s (e_d + e_u) I[0, 4] + 
 6 (97 e_d + 2320 e_s + 97 e_u) I[0, 5] - 
 171520 m_c m_s (e_d \nonumber\\
 &+ e_u) I[1, 3] - 
 4 e_c (5559 I[0, 5] + 4480 m_c m_s (8 I[0, 4] + 9 I[1, 3])) + 
 5 (-9240 e_c + 1865 e_d - 1312 e_s \nonumber\\
 &+ 1865 e_u) I[1, 4]\Big]\nonumber\\
 & +\frac{C_2\, m_c\,m_s}{2^{20}\times 3^4 \times 5^2   \pi^5} \Big[ 9 (e_d + e_u) (I[0, 5] + 35 I[1, 4]) - 16 e_c (7 I[0, 5] + 95 I[1, 4])\Big] \nonumber\\
 &+
 \frac{C_3\, m_c\,m_s}{2^{21}\times 3^4 \times 5^2   \pi^5} \Big[ 3 (-11 e_c + 6 e_d + 6 e_u) I[0, 5] + 
 5 (152 e_c - 21 e_d + 21 e_u) I[1, 4]\Big],
\end{align}
 
 \subsection*{3-The obtained sum rule for the magnetic dipole moment
of the $J_{\mu}^3$ interpolating current}
  \begin{align}
      \rho_3 (\mathrm{M^2},\mathrm{s_0}) &=   F_1^{J^3_{\mu}}(\mathrm{M^2},\mathrm{s_0}) - \frac{1}{m_{J_\mu^3}} F_2^{J^3_{\mu}}(\mathrm{M^2},\mathrm{s_0}),
 \end{align}
 where, 
 \begin{align}
 F_1^{J^3_{\mu}}(\mathrm{M^2},\mathrm{s_0})&= -  \frac{1}{2^{23}\times 3 \times 5^2 \times 7^2 \pi^7} \Big[  49 m_c m_s (e_d + e_u)  I[0, 6] + 13 (e_d + e_s + e_u) I[0, 7]  \Big]
 \nonumber\\
 &+ \frac{C_1 C_2 C_3}{2^{19}\times 3^7 \times 5  \pi^3} \Big[ 77 (5 e_c + 4 e_d +4 e_u ) I[0, 2]  \Big]
 \nonumber\\
 &- \frac{11 C_1 C_2^2}{2^{19}\times 3^7   \pi^3} \Big[ (e_c - 2 e_s) I[0, 2]\Big]
 \nonumber\\
 &- \frac{ C_1 C_2 m_0^2}{2^{22}\times 3^5 \pi^5}  \Big[ m_c(269 e_c + 19 e_d + 2 e_s + 19 e_u)  I[0, 2]  \Big]
   \nonumber
  \end{align}
  \begin{align}
 &- \frac{ C_1 C_3 m_0^2}{2^{23}\times 3^5 \times 5  \pi^5}\Big[ \Big(4 (e_d + e_u) (25 m_c + 11 m_s) + e_c (1750 m_c + 99 m_s)\Big) I[0, 2  \Big]
  \nonumber\\
 &+ \frac{ C_1 C_2}{2^{22}\times 3^7 \times 5  \pi^5}\Big[ \Big (3 m_c (3518 e_c + 169 e_d + 674 e_s + 169 e_u) + 2476 e_c m_s - 
   30 m_s (e_d + e_u)\Big) I[0, 3] \Big]
  \nonumber\\
 &+ \frac{ C_1 C_3}{2^{23}\times 3^7 \times 5  \pi^5}\Big[ \Big (2 e_c (7176 m_c - 1375 m_s) + (e_d + 
      e_u) (960 m_c + 1081 m_s)\Big) I[0, 3] \Big]
 \nonumber\\
 &- \frac{e_c C_2 C_3}{2^{17}\times 3^3 \times 5  \pi^3}I[0, 4]
 \nonumber\\
 &+ \frac{e_c C_2^2}{2^{18}\times 3^4 \times 5  \pi^3}\Big[352 m_c m_s I[0, 3] - 3 I[0, 4] \Big]
 \nonumber\\
 &+ \frac{ C_1}{2^{28}\times 3^6 \times 5^2  \times 7 \pi^7}\Big[  -840 m_c m_s (488 e_c + 29 e_d + 29 e_u)  I[0, 
   4] + (-38892 e_c + 23435 e_d + 57554 e_s \nonumber\\
   &+ 23435 e_u) I[0, 5]  \Big]
  \nonumber\\
 &+ \frac{  C_2}{2^{20}\times 3^2 \times 5^2  \pi^5}\Big[  (7 m_c (e_d + 2 e_s + e_u) + 8 e_c m_s) I[0, 5]  \Big]
\nonumber\\
 &+ \frac{  C_3}{2^{20}\times 3^2 \times 5^2 \times 7  \pi^5}\Big[ (e_d + e_u) (49 m_c - 12 m_s) I[0, 5]   \Big],\\
 \nonumber\\
 F_2^{J^3_{\mu}}(\mathrm{M^2},\mathrm{s_0})&= \frac{ m_c}{2^{25}\times 3^5 \times 5^3 \times 7^2 \pi^7} \Big [ 16 \Big (3430 m_c m_s (e_d + e_u)  I[0, 6] - 
    594 (e_d + e_s + e_u) I[0, 7] + 
    8085  m_c m_s (e_d \nonumber\\
    &+ e_u) I[1, 5]\Big) - 
 46116 (e_d + e_s + e_u) I[1, 6] + 
 7 e_c \Big (-40 m_c m_s (107 I[0, 6] + 492 I[1, 5]) + 
    27 (41 I[0, 7] \nonumber\\
    &+ 382 I[1, 6])\Big)  \Big]\nonumber\\
 &+\frac{C_1 C_2 C_3 \, m_c}{2^{17}\times 3^6  \pi^3} \Big[2 m_c m_s (-173 e_c + 13 e_d + 13 e_u) I[0, 1] + 
 36 (4 e_c + e_d + e_u) I[0, 2] - 9 (17 e_c + e_d \nonumber\\
    &+ e_u) I[1, 1]
     \Big]\nonumber\\
 &+\frac{ C_1 C_2^2 \, m_c}{2^{16}\times 3^4  \pi^3}\Big[ (3 e_c + e_s)I[0, 2] - 3 e_c I[1, 1] \Big]\nonumber\\
 &+ \frac{C_1 C_2 m_0^2\,m_c}{2^{22}\times 3^5  \pi^5}
 \Big[  \Big (736 e_c m_c - 21 m_c (5 e_d + 13 e_s + 5 e_u) + 
    72 m_s (4 e_c + e_d + e_u) \Big) I[0, 
   2] - \Big (m_c (1346 e_c + e_d \nonumber\\
    &- 52 e_s + e_u) + 
    18 m_s (17 e_c + e_d + e_u) \Big) I[1, 1] \Big]
   \nonumber\\
    &+ \frac{C_1 C_3 m_0^2\,m_c}{2^{23}\times 3^6   \pi^5}
 \Big[  -3 (e_d + e_u) \Big ((93 m_c - 8 m_s) I[0, 2] + 
    2 (91 m_c + 4 m_s) I[1, 1]\Big) + 
 4 e_c \Big (3 (139 m_c - 47 m_s)   \nonumber\\
    & \times I[0, 2]+ 
    4 (-249 m_c + 26 m_s) I[1, 1]\Big)\Big]
 \nonumber\\
 &+ \frac{C_1 C_2\,m_c}{2^{21}\times 3^7\pi^5}
 \Big[ \Big (m_c (157 e_c + 337 e_d + 692 e_s + 337 e_u) - 
    18 m_s (7 e_c + 13 e_d + 13 e_u) \Big) I[0, 3] + 
 3 \Big (-m_c (\nonumber\\
    & \times -1720 e_c + e_d + 218 e_s + e_u) + 
    36 m_s (17 e_c + e_d + e_u) \Big) I[1, 2] \Big]
 \nonumber\\
 &+ \frac{C_1 C_3 \,m_c}{2^{23}\times 3^7 \times 5   \pi^5}
 \Big[ 2 \Big ((e_d + e_u) (3190 m_c + 481 m_s) + 
    e_c (4790 m_c + 964 m_s)\Big) I[0, 3] + 
 3 \Big (4 e_c (4210 m_c \nonumber\\
    & - 957 m_s) + (e_d + 
       e_u) (1960 m_c - 477 m_s)\Big) I[1, 2]  \Big]
 \nonumber\\
 & +\frac{C_2 C_3 \,m_c^2 m_s}{2^{14}\times 3^5 \times 5 \pi^5} \Big[  -20 e_c (I[0, 3] - 12 I[1, 2]) + 11 (e_d + e_u) (I[0, 3] - 3 I[1, 2]) \Big]
  \nonumber\\
 & +\frac{C_1\, m_c}{2^{30}\times 3^6 \times 5^2 \pi^7} \Big[ -160 m_c m_s (2059 e_c + 666 e_d + 666 e_u)  I[0, 4] - 
 2 (10806 e_c + 1499 e_d + 5360 e_s  \nonumber\\
    &+ 1499 e_u) I[0, 5] - 
 1280 m_c m_s (599 e_c + 56 (e_d + e_u))  I[1, 3] + 
 15 (1552 e_c - 2363 e_d - 9200 e_s  \nonumber\\
    &- 2363 e_u) I[1, 4]\Big]\nonumber\\
 & +\frac{C_2\, m_c^2}{2^{19}\times 3^4 \times 5^2   \pi^5} \Big[  4 e_c (7 I[0, 5] + 45 I[1, 4]) - (e_d + 2 e_s + e_u) (19 I[0, 5] + 
    65 I[1, 4])  \Big] \nonumber
    \end{align}
    
  \begin{align}
 &+
 \frac{C_3\, m_c}{2^{21}\times 3^4 \times 5^2   \pi^5} \Big[ \Big (56 e_c m_c - 76 m_c (e_d + e_u) - 
    27 m_s (5 e_c - 4 e_d - 4 e_u)\Big)  I[0, 5] + 
 10 \Big (9 e_c (4 m_c + 21 m_s)  \nonumber\\
    &- (e_d + 
       e_u) (26 m_c + 63 m_s)\Big) I[1, 4]\Big],
\end{align}
 \noindent  where $C_1 =\langle g_s^2 G^2\rangle$ is gluon condensate;  $C_2 =\langle \bar q q \rangle$ and $C_3 =\langle \bar ss \rangle$ stand for u/d-quark and s-quark condensates, respectively. 
 The function $\mathrm{I}[n,m]$ is given as 
\begin{align}
 \mathrm{I}[n,m]&= \int_{\mathcal M}^{\mathrm{s_0}} ds~ e^{-s/\mathrm{M^2}}~
 s^n\,(s-\mathcal M)^m,\nonumber
 \end{align}
\noindent where $\mathcal{M}= (2m_c+m_s)^2$.

\bibliographystyle{elsarticle-num}
\bibliography{PcsMM.bib}
\end{document}